\documentclass[tradiabstract]{aa} 
\usepackage{graphicx}
\usepackage{txfonts}
\usepackage{natbib}
\usepackage{longtable}
\usepackage{lscape}
\usepackage{multirow}
\usepackage[colorlinks=true, allcolors=blue]{hyperref}

\bibpunct{(}{)}{;}{a}{}{,}

\newcommand{\kms}{km\,s$^{-1}$}

\newcommand{\degree}{$^{\circ}$}

\begin{document}
\title{Searching for further evidence for cloud-cloud collisions in L1188}

\author{Y. Gong\inst{1,2}, X. D. Tang\inst{3,1}, C. Henkel\inst{1,4,3}, K. M. Menten\inst{1}, R.~Q. Mao\inst{2}, Y. Wang\inst{5}, M.-Y. Lee\inst{6}, W.~S. Zhu\inst{7}, Y. Lin\inst{1}, S.~B. Zhang\inst{2}, X.~P. Chen\inst{2}, W.~J. Yang\inst{2}}

\offprints{Y. Gong, \email{ygong@mpifr-bonn.mpg.de, gongyan2444@gmail.com}}

\institute{
  Max-Planck-Institut f{\"u}r Radioastronomie, Auf dem H{\"u}gel 69, D-53121 Bonn, Germany \\e-mail: ygong@mpifr-bonn.mpg.de, gongyan2444@gmail.com
  \and
  Purple Mountain Observatory \& Key Laboratory for Radio Astronomy, Chinese Academy of Sciences, 10 Yuanhua Road, 210033 Nanjing, PR China
  \and
  Xinjiang Astronomical Observatory, Chinese Academy of Sciences, 830011 Urumqi, PR China
  \and
  Astronomy Department, Faculty of Science, King Abdulaziz University, P.O. Box 80203, Jeddah 21589, Saudi Arabia
  \and
  Max-Planck Institute for Astronomy, K{\"o}nigstuhl 17, 69117 Heidelberg, Germany
  \and
  Korea Astronomy and Space Science Institute 776, Daedeokdae-ro, Yuseong-gu, Daejeon, 34055, Republic of Korea
  \and
  School of Physics and Astronomy, Sun Yat-Sen University, Zhuhai 519082, People's Republic of China
}

\date{Received date ; accepted date}

\abstract{In order to search for further observational evidence of cloud-cloud collisions in one of the promising candidates, L1188, we carried out observations of multiple molecular lines toward the intersection region of the two nearly orthogonal filamentary molecular clouds in L1188. Based on these observations, we find two parallel filamentary structures, both of which have at least two velocity components being connected with broad bridging features. We also found a spatially complementary distribution between the two molecular clouds, as well as enhanced $^{13}$CO emission and $^{12}$CO self-absorption toward their abutting regions. At the most blueshifted velocities, we unveil a 1~pc-long arc ubiquitously showing $^{12}$CO line wings. We discover two 22 GHz water masers, which are the first maser detections in L1188. An analysis of line ratios at a linear resolution of 0.2 pc suggests that L1188 is characterised by kinetic temperatures of 13--23~K and H$_{2}$ number densities of 10$^{3}$--10$^{3.6}$ cm$^{-3}$. On the basis of previous theoretical predictions and simulations, we suggest that these observational features can be naturally explained by the scenario of a cloud-cloud collision in L1188, although an additional contribution of stellar feedback from low-mass young stellar objects cannot be ruled out.}    
\titlerunning{Searching for further evidence for cloud-cloud collisions in L1188}
\authorrunning{Y. Gong et al.}
\maketitle

\section{Introduction}\label{sec:info}
Cloud-cloud collisions have been proposed to play an important role in molecular cloud evolution, star formation, and, on a grander scale, in galaxy evolution \citep[e.g.,][]{2007ARA&A..45..565M}. Theoretically, many studies suggest that the formation of giant molecular clouds could be due to agglomeration via successive inelastic cloud-cloud collisions \citep[e.g.,][]{1979ApJ...229..567K,1987ApJ...314...10R}. On the other hand, cloud-cloud collisions are thought to be important for star formation. Previous studies suggested that a cloud-cloud collision could induce star-forming seeds in the shock-compressed interface \citep[e.g.,][]{1992PASJ...44..203H,2010MNRAS.405.1431A}. Such a process would favor the formation of massive molecular filaments \citep{2018PASJ...70S..53I} from which massive stars can fragment \citep{2013ApJ...774L..31I}. Numerical simulations suggest that such a mechanism greatly enhances star formation rates and efficiencies \citep{2017ApJ...841...88W}. Furthermore, cloud-cloud collisions may make significant contributions to global star formation rates in spiral galaxies because such collisions are expected to occur frequently \citep[once every 8--10 Myr in spiral arms,][]{2015MNRAS.446.3608D}. In addition, \citet{2017MNRAS.471.2002L} argued that kinematic dissipation by cloud-cloud collisions might lead to the formation of giant massive clumpy structures in high-redshift gas-rich galaxies. Identifying cloud-cloud collisions is indispensable in order to verify these theories, while it is still challenging.

From an observational point of view, a few nearby molecular clouds are believed to exhibit features of cloud-cloud collisions \citep{2011A&A...528A..50D,2012ApJ...746...25N}. These candidates are mainly identified by blueshifted and redshifted velocity fields. Toward the rather distant infrared dark cloud G035.39$-$00.33 ($\sim$2.9 kpc), parsec-scale SiO emission \citep[e.g.,][]{2010MNRAS.406..187J} and the spatio-kinematic offset between different tracers \citep[e.g.,][]{2013MNRAS.428.3425H,2018MNRAS.478L..54B} are also proposed to result from cloud-cloud collisions. Numerical simulations have shown that broad bridge features in position-velocity diagrams can serve as an isolated signature of cloud-cloud collisions \citep{2015MNRAS.454.1634H,2015MNRAS.450...10H}. These criteria have already led to many cloud-cloud collision candidates, such as NGC 3603 \citep{2014ApJ...780...36F}, RCW 38 \citep{2016ApJ...820...26F}, M20 \citep{2011ApJ...738...46T,2017ApJ...835..142T}, L1188 \citep{2017ApJ...835L..14G}, RCW 79 \citep{2018PASJ...70S..45O}, NGC 6618/M17 \citep{2018PASJ...70S..42N}, GM24 \citep{2018PASJ...70S..44F}, NGC 6334/NGC 6357 \citep{2018PASJ...70S..41F}, and the references therein. However, there are still some uncertainties as to whether these candidates are true collisions or not; because cloud morphologies and internal motions are usually complex and a competitive process, the feedback by massive stars, complicates the picture. Therefore, additional observational features of cloud-cloud collisions is beneficial for robust identifications.  

\begin{figure}[!htbp]
\centering
\includegraphics[width = 0.45 \textwidth]{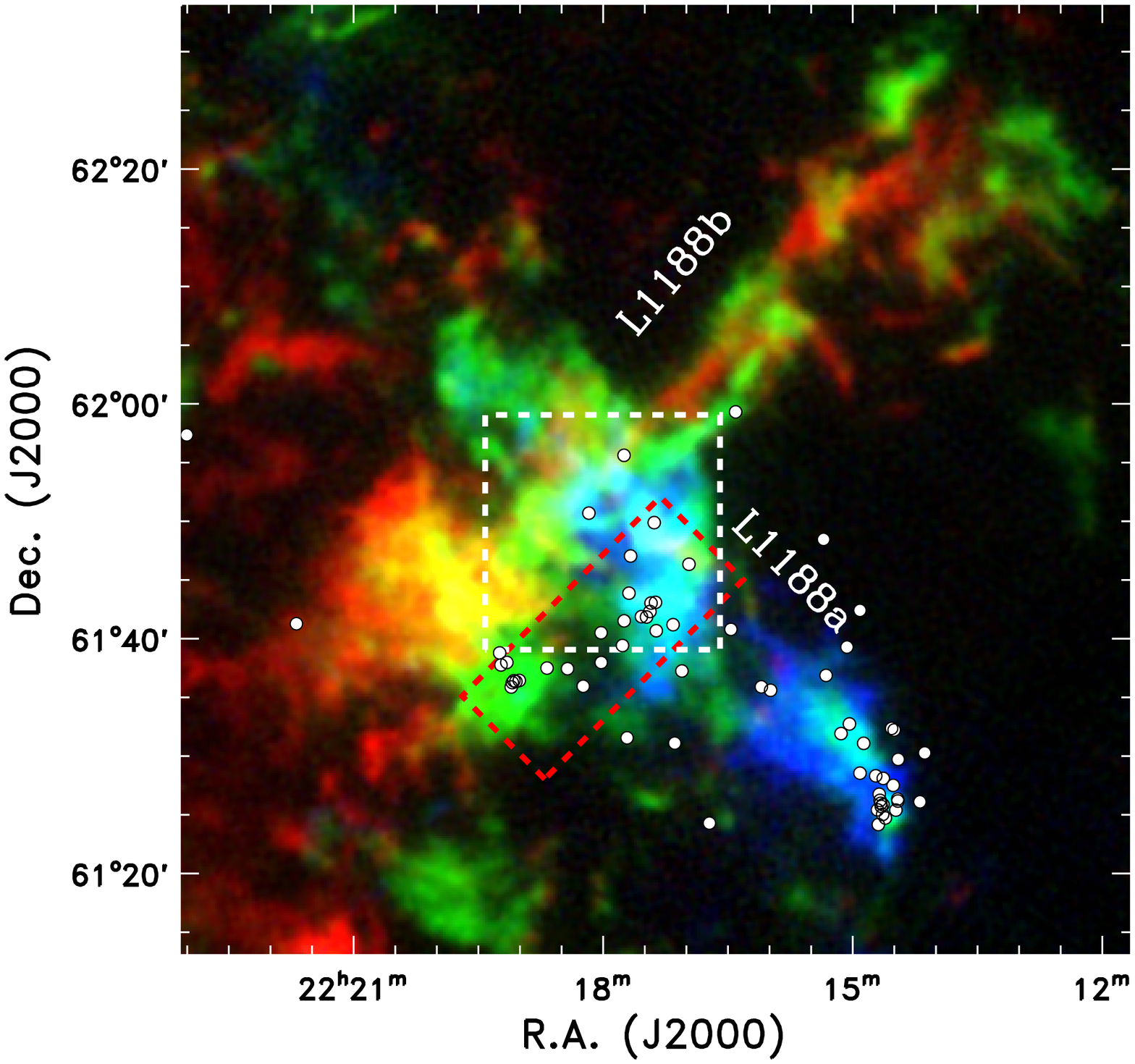}
\caption{{Overview of L1188, which zooms in on Fig.~3a of \citet{2017ApJ...835L..14G} (red: $^{12}$CO (1--0) intensity map integrated between $-$8.0 and $-$5.0 \kms; green: $^{12}$CO (1--0) intensity map integrated between $-$11 and $-$8.0~\kms; blue: $^{12}$CO (1--0) intensity map integrated between $-$14 and $-$11~\kms.). The observed region is indicated by the white dashed box. The white circles represent the young stellar object (YSO) candidates associated with L1188, which are aligned in a filamentary structure inside the red dashed box \citep{2019MNRAS.484.1800S}.}\label{Fig:overview}}
\end{figure}

L1188, located at a distance of $\sim$800 pc \citep{2017ApJ...835L..14G,2019MNRAS.484.1800S}, was first mapped in $^{13}$CO (1--0) with the Nagoya-4 m telescope \citep{1995A&A...300..525A}. A subsequent study suggests that L1188 is one of the most promising cloud-cloud collision candidates with a total molecular gas mass of 3.9$\times 10^{3}$~M$_{\odot}$ and without massive stars ($M>8M_{\odot}$) inside \citep{2017ApJ...835L..14G}. L1188 is thus much less affected by the energetic feedback from massive stars when compared with the aforementioned candidates associated with massive stars. Furthermore, previous CO observations demonstrate that L1188 has a relatively simple structure, which consists of two nearly orthogonal filamentary molecular clouds, namely L1188a and L1188b \citep{2017ApJ...835L..14G}. These properties allow the assignation of observational features to different phenomena. In order to search for further evidence of cloud-cloud collisions, we performed a multiple molecular line study toward the intersection region in L1188, covering a size of 20\arcmin$\times$20\arcmin (4.7 pc$\times$4.7 pc, see Fig.~\ref{Fig:overview}).


\section{Observations and data reduction}\label{Sec:obs}
We performed mapping observations of five different molecules with four different telescopes. Table~\ref{Tab:lin} summarizes these observations.

\subsection{PMO-13.7 m observations}
The intersection region was observed in HCO$^{+}$ (1--0),  SiO (2--1) and CH$_{3}$OH ($8_{0}$--$7_{1}$ A$^{+}$) with the Purple Mountain Observatory 13.7 m telescope (PMO-13.7 m) from 2018 May 1 to 15 (project code: 18A003). We employed the 3$\times$3 beam sideband separation Superconducting Spectroscopic Array Receiver (SSAR) and fast Fourier transform spectrometers (FFTSs) for our observations \citep{2012ITTST...2..593S}. Each of the 18 FFTS modules provides 16384 channels, covering an instantaneous bandwidth of 1 GHz. This results in a channel spacing of 61~kHz, that is, 0.20~\kms\,at 86~GHz. The region was observed in the on-the-fly mode with a scanning speed of 50\arcsec\,per second and a dump time of 0.3 seconds. These observations took around 40 hours in total.

The antenna temperature, $T_{\rm A}^{*}$, was obtained with the standard chopper wheel method \citep{1976ApJS...30..247U}. Throughout this work, the antenna temperature is converted into the main beam brightness temperature scale, $T_{\rm mb}$, with the relation, $T_{\rm mb}=\frac{T_{\rm A}^{*}}{\eta_{\rm mb}}$, where $\eta_{\rm mb}$ is the main beam efficiency. The main beam efficiency is 57\% at 86~GHz according to the telescope's status report \footnote{http://www.radioast.nsdc.cn/ztbg/ztbg2015-2016engV2.pdf}. The flux calibration error is estimated to be roughly 10\%. During the observations, system temperatures were 145--301 K on a $T_{\rm A}^{*}$ scale. The half-power beam width (HPBW) is about 60\arcsec\,at the observed frequencies. The pointing was found to have an uncertainty of 5\arcsec, which is less than 1/10 of the HPBW. 

The GILDAS\footnote{https://www.iram.fr/IRAMFR/GILDAS/} software was used to reduce the data \citep{2005sf2a.conf..721P}. Raw data were gridded into regular separations of 30\arcsec\,between two adjacent pixels. The achieved sensitivities are $\sim$0.08~K on a $T_{\rm mb}$ scale for SiO (2--1) at a channel width of 0.20~\kms. 

CO (1--0) and $^{13}$CO (1--0) data, also observed with the PMO-13.7 m telescope as part of the Milky Way Imaging Scroll Painting \citep[MWISP\footnote{http://www.radioast.nsdc.cn/mwisp.php};][]{2019ApJS..240....9S} project, have already been presented in \citet{2017ApJ...835L..14G} and are also used in our analysis. 

\subsection{IRAM-30 m observations}
CO (2--1) observations (project code: 138-16, 023-18) were performed with the Institut de Radioastronomie Millim{\'e}trique 30 m (IRAM-30 m) telescope and the 9 dual-polarization pixel HEterodyne Receiver Array \citep[HERA,][]{2004A&A...423.1171S} in 2017 March and 2018 August. The selected region was observed in the on-the-fly mode, scanning along right ascension and declination directions. The dump time is typically 0.5 seconds with a scanning speed of 6\arcsec\, per second. The calibration was done every 6--8 minutes. FFTSs were used as backends to record signals from both linear polarizations. These FFTSs provide a channel spacing of 195~kHz, corresponding to a channel width of 0.25~\kms\,at 230~GHz. The focus was adjusted via observations of the radio galaxy 2021+614. Pointing, which was regularly checked on nearby strong continuum sources, was found to have a typical rms uncertainty of $\sim$5\arcsec. The spectral data are expressed in units of main beam brightness temperature by adopting a forward efficiency of 92\%\,and a main beam efficiency of 57\%. The HBPW is about 11\arcsec\,at 230 GHz. The typical noise is about 0.5~K at a channel width of 0.25~\kms. These IRAM-30 m data were also analyzed with the GILDAS software. 
\subsection{JCMT-15 m observations}
We carried out $^{12}$CO (3--2) and $^{13}$CO (3--2) observations (project code: M17AP002) toward the central 20\arcmin$\times$20\arcmin\,region in L1188 with the James Clerk Maxwell Telescope (JCMT) from 2017 May 31 to June 28. The 16-element Heterodyne Array Receiver Program (HARP) was used as front-end, while the Auto-Correlation Spectral Imaging System (ACSIS) was employed as back-end \citep{2009MNRAS.399.1026B}. ACSIS provides 7172 channels, each of which is 30.5 kHz-wide. This is equivalent to a width of 0.026~\kms\, at 345 GHz. We used the on-the-fly mode to scan the observed region in orthogonal directions. At 345 GHz, the HPBW is about 14$\rlap{.}^{\prime\prime}$5 and the main beam efficiency is 61\%. Pointing was checked using JCMT standard sources, resulting in a typical rms uncertainty of $\sim$3\arcsec. 

The observed data were reduced with the ORAC Data Reduction pipeline \citep[ORAC-DR,][]{2015MNRAS.453...73J} of the Starlink\footnote{http://starlink.eao.hawaii.edu/starlink} software \citep{2014ASPC..485..391C}. For the following analysis, all spectra have been binned to a channel width of 0.27~\kms\,to improve signal-to-noise ratios. The typical noise levels are about 1~K and 0.7~K at a channel width of 0.27~\kms for $^{12}$CO (3--2) and $^{13}$CO (3--2), respectively. Velocities are given with respect to the local standard of rest (LSR) throughout this work.
\subsection{Effelsberg-100 m observations}
We carried out H$_{2}$O (6$_{1,6}$--5$_{2,3}$) measurements (project code: 22-16, 59-17) in a position-switching mode with the double-beam and dual-polarization K-band receiver of the 100-m telescope at Effelsberg/Germany\footnote{The 100-m telescope at Effelsberg is operated by the Max-Planck-Institut f{\"u}r Radioastronomie (MPIFR) on behalf of the Max-Planck Gesellschaft (MPG).} during 2019 April. The two beams are at the same declination but separated by 192\arcsec\, in right ascension. We performed a raster map toward the region showing $^{12}$CO wing emission (see results given below) with a grid spacing of 15\arcsec, leading to a coverage of 360\arcsec$\times$75\arcsec. The FFTS provides a bandwidth of 300 MHz, consisting of 65536 channels with a channel spacing of 4.6 kHz which is equivalent to 0.06~\kms\, at 22 GHz. The full width at half maximum (FWHM) beamsizes are about 40\arcsec\,at 22 GHz. NGC 7027 was used for pointing, focus, and flux calibration, and the flux calibration accuracy is estimated to be $\pm$20\%. Nearby pointing results were obtained toward the radio galaxy 2021+614, and the rms pointing uncertainty was found to be 5\arcsec. Typical system temperatures on a $T_{\rm A}^{*}$ scale are 49--80 K. The main beam efficiency was 59.7\%. The conversion factor from flux density ($S_{\nu}$) to main beam temperature ($T_{\rm mb}$) is $T_{\rm mb}/S_{\nu}$=1.75 K/Jy at 22 GHz. These Effelsberg-100 m data were also analyzed with the GILDAS software. 



\begin{table*}[!hbt]
\caption{Observed parameters related to the molecular lines presented in this work.}\label{Tab:lin}
\normalsize
\centering
\begin{tabular}{ccccccccc}
\hline \hline
line             & Frequency         & $E_{\rm u}/k$      & $\theta_{\rm beam}$    & $\delta \varv$     & $\sigma$ & mapping area             & telescope  \\
                 & (GHz)             & (K)              &  (\arcsec)           & (\kms)             &   (K)    & (\arcsec$\times$\arcsec) &            \\
(1)              & (2)               & (3)              & (4)                  & (5)                &  (6)     & (7)                         &  (8)       \\
\hline
$^{12}$CO $J=1-0$ & 115.27120           & 6   & 52   & 0.16 & 0.5           & 1200$\times$1200    & PMO-13.7 m       \\
$^{13}$CO $J=1-0$ & 110.20135           & 5   & 55   & 0.17 & 0.3           & 1200$\times$1200    & PMO-13.7 m       \\
$^{12}$CO $J=2-1$ & 230.53800           & 17  & 11   & 0.25 & 0.5           & 1200$\times$1200    & IRAM-30 m        \\
$^{12}$CO $J=3-2$ & 345.79599           & 33  & 14.5 & 0.27 & 1.0           & 1200$\times$1200    & JCMT-15 m        \\
$^{13}$CO $J=3-2$ & 330.58797           & 32  & 15.1 & 0.27 & 0.7           & 1200$\times$1200    & JCMT-15 m        \\
HCO$^{+}$ $J=1-0$ & 89.18852            & 4   & 61   & 0.20  & 0.08         & 1200$\times$1200    & PMO-13.7 m       \\
SiO $J=2-1$      & 86.84696            & 6   & 61   & 0.20  & 0.08          & 1200$\times$1200   & PMO-13.7 m       \\
CH$_{3}$OH $J_{K}=8_{0}-7_{1} A^{+}$ & 95.16939 & 84  & 55   & 0.18 & 0.08 & 1200$\times$1200      & PMO-13.7 m       \\
H$_{2}$O $J_{K_a,K_c}=6_{1,6}-5_{2,3}$  &  22.23508 & 644 & 40   & 0.06 &  0.2         & 360$\times$75     & Effelsberg-100 m \\
\hline
\end{tabular}
\tablefoot{(1) Observed transition. (2) Rest frequency. (3) The upper energy of the observed transition. (4) The beam size. (5) The channel width in units of \kms. (6) The rms noise level. (7) The mapped region. (8) The telescope.}
\normalsize
\end{table*}

\section{Results}\label{Sec:res}
\subsection{Molecular distribution and kinematics}\label{res.mor}
\begin{figure*}[!htbp]
\centering
\includegraphics[width = 0.95 \textwidth]{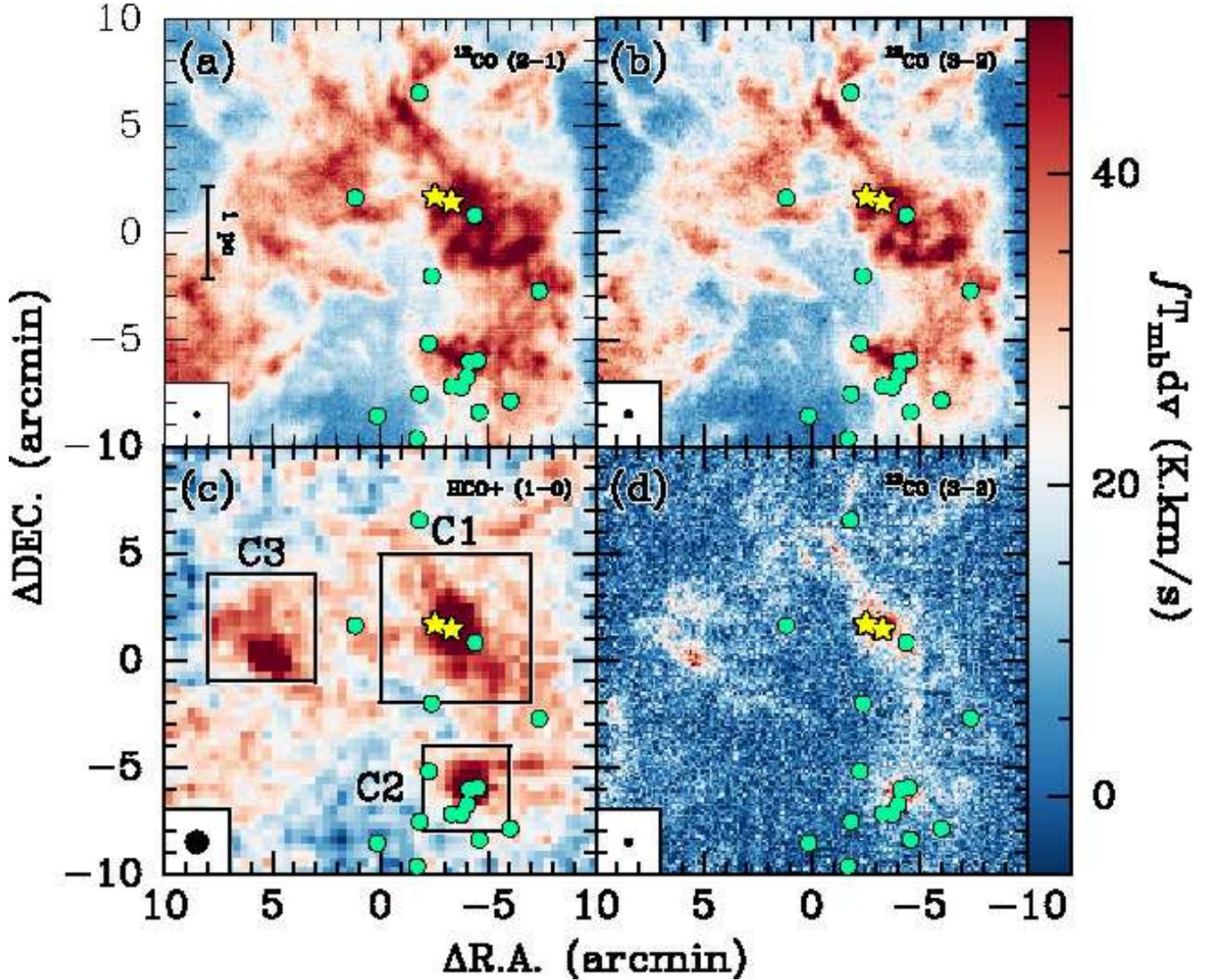}
\caption{{Maps of main beam brightness temperature integrated from $-$14 to $-$5~\kms for (a) $^{12}$CO (2--1), (b) $^{12}$CO (3--2), (c) HCO$^{+}$ (1--0), and (d) $^{13}$CO (3--2). The color bar represents the integrated intensities in units of K~\kms. The integrated intensities have been scaled by a factor of 4/3, 40, and 5 for $^{12}$CO (3--2), HCO$^{+}$ (1--0), and $^{13}$CO (3--2), respectively, in order to match the color bar. The three clumpy regions are designated as C1, C2, and C3 in Fig.~\ref{Fig:m0}c. In each panel, the green filled circles represent the YSO candidates from \citet{2019MNRAS.484.1800S} and the two yellow pentagrams represent the two 22 GHz water maser positions discussed in Sect.~\ref{sec.maser}. The beam size is shown in the lower left corner of each panel. The (0, 0) offset corresponds to $\alpha_{\rm J2000}$=22$^{\rm h}$18$^{\rm m}$00$\rlap{.}^{\rm s}$44, $\delta_{\rm J2000}$=61\degree49\arcmin03$\rlap{.}^{\prime\prime}$7.}\label{Fig:m0}}
\end{figure*}

Our new observations have led to the detection of extended emission in $^{12}$CO (2--1), $^{12}$CO (3--2), HCO$^{+}$ (1--0), and compact filamentary structures in $^{13}$CO (3--2). The integrated intensity maps are shown in Fig.~\ref{Fig:m0}. The spatial distributions in $^{12}$CO (2--1) and $^{12}$CO (3--2) are similar to $^{12}$CO (1--0), but both have a factor of $\sim$4 higher angular resolutions compared with previous $^{12}$CO (1--0) observations \citep[e.g.,][]{2017ApJ...835L..14G}. Previously, only five positions had been observed in HCO$^{+}$ (1--0) \citep{2013AN....334..920V}, so the distribution of HCO$^{+}$ (1--0) emission is revealed for the first time by our measurements. We designate the three clumpy regions, which are indicated in Fig.~\ref{Fig:m0}c, as C1, C2, and C3. The $^{13}$CO (3--2) data have rather low signal-to-noise ratios, so their resolution has been smoothed to 55\arcsec\, to achieve a higher dynamic range hereafter. It turns out that the YSO candidates appear to cluster toward region C2 where strong and clumpy HCO$^{+}$ (1--0) and $^{13}$CO (3--2) emission is present. In contrast, only one YSO candidate is associated with region C1, while no YSO candidate is found in region C3. This may indicate that region C2 is the most evolved region, followed by region C1 and C3.  



\begin{figure*}[!htbp]
\centering
\includegraphics[width = 0.95 \textwidth]{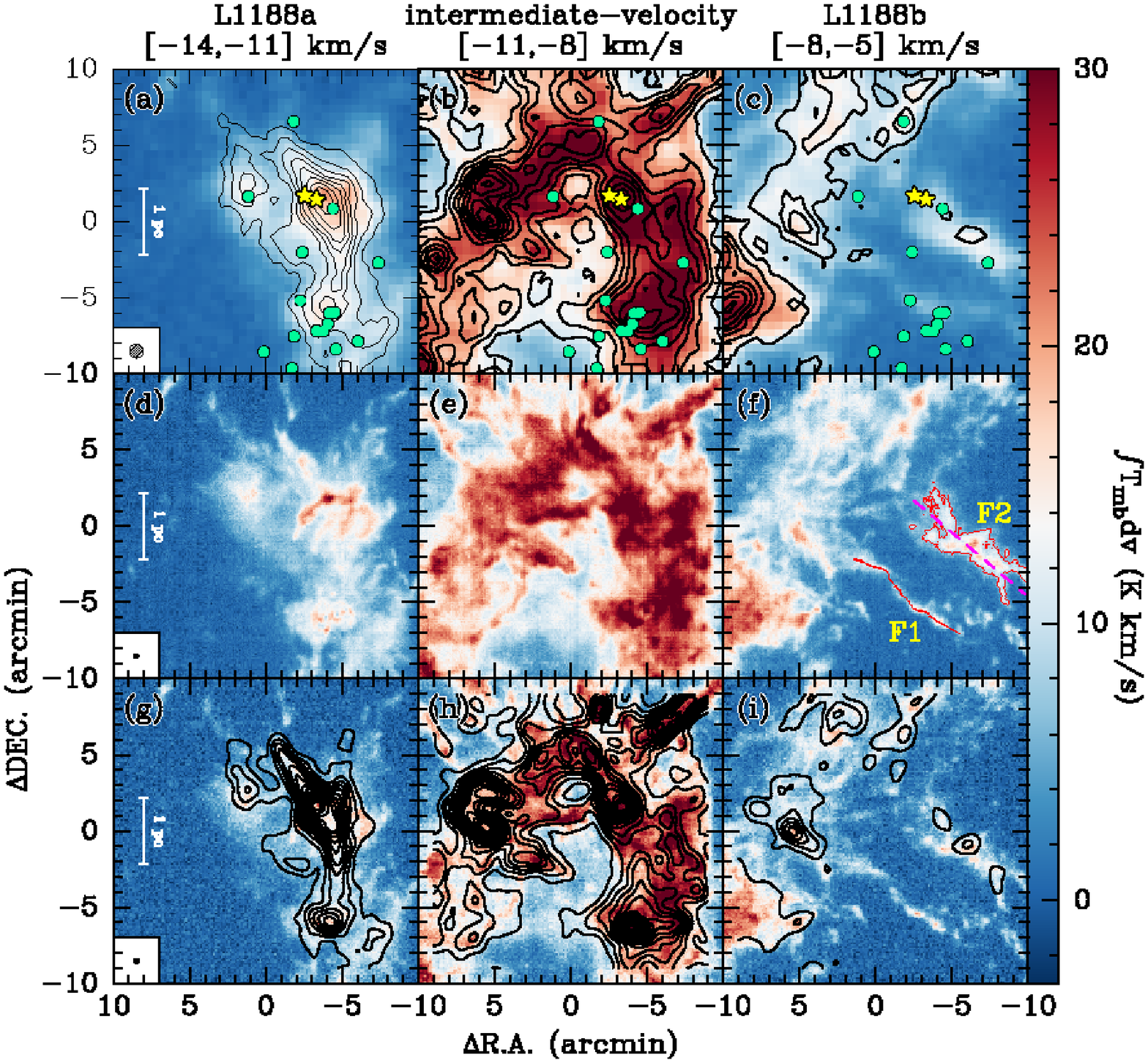}
\caption{{$^{12}$CO (1--0) intensity map (color scale) integrated (a) from $-$14 to $-$11~\kms, (b) from $-$11 to $-$8~\kms, (c) from $-$8 to $-$5~\kms. In Figures~\ref{Fig:mor}a--c, the contours represent the integrated intensity maps of $^{13}$CO (1--0), and their integrated velocity ranges are the same as those of the corresponding CO maps. The contours start from 0.9~K~\kms\,(5$\sigma$), and increase by 0.9~K~\kms. Figures~\ref{Fig:mor}d--f and Figures~\ref{Fig:mor}g--i are similar to Figures~\ref{Fig:mor}a--c but for $^{12}$CO (2--1) and $^{12}$CO (3--2), respectively. In Figures~\ref{Fig:mor}g--i, the contours represent the smoothed (55\arcsec) integrated intensity maps of $^{13}$CO (3--2), and start from 0.3 K~\kms\,(5$\sigma$), and increase by 0.3~K~\kms. Figures~\ref{Fig:mor}g--i have been scaled by a factor of two to match the color bar. In Figs.~\ref{Fig:mor}a--\ref{Fig:mor}c, the green filled circles represent the YSO candidates from \citet{2019MNRAS.484.1800S} and the two yellow pentagrams represent the two 22 GHz water maser positions discussed in Sect.~\ref{sec.maser}. The filamentary structures F1 and F2 are outlined in Fig.~\ref{Fig:mor}f. The beam size for CO (1--0), CO (2--1), and CO (3--2) are shown in the lower left corner of Figs.~\ref{Fig:mor}a, \ref{Fig:mor}d, and \ref{Fig:mor}g, respectively. In all panels, the (0, 0) offset corresponds to $\alpha_{\rm J2000}$=22$^{\rm h}$18$^{\rm m}$00$\rlap{.}^{\rm s}$44, $\delta_{\rm J2000}$=61\degree49\arcmin03$\rlap{.}^{\prime\prime}$7. }\label{Fig:mor}}
\end{figure*}

Figure~\ref{Fig:mor} presents the intensity maps of $^{12}$CO (1--0), $^{13}$CO (1--0) (in contours), $^{12}$CO (2--1), $^{12}$CO (3--2), and $^{13}$CO (3--2) (in contours) with different integrated velocity ranges. In this work, we refer to the emission integrated from $-$14 to $-$11~\kms\, as L1188a (see Figs.~\ref{Fig:mor}a, \ref{Fig:mor}d, and \ref{Fig:mor}g), the emission integrated from $-$11 to $-$8~\kms\, as the intermediate-velocity region (see Figs.~\ref{Fig:mor}b, \ref{Fig:mor}e, and \ref{Fig:mor}h), and the emission integrated from $-$8 to $-$5~\kms\,as L1188b (see Figs.~\ref{Fig:mor}c, \ref{Fig:mor}f, and \ref{Fig:mor}i). Thanks to the achieved high angular resolutions of 12--15\arcsec, $^{12}$CO (2--1) and $^{12}$CO (3--2) intensity maps show more fluffy structures than our previous $^{12}$CO (1--0) maps. Short wispy and filamentary structures are ubiquitously present at the edges of L1188a, the intermediate-velocity region, and L1188b (see Figs.~\ref{Fig:mor}d--\ref{Fig:mor}i); also, they appear to be randomly oriented. In the intermediate-velocity region, cavity-like structures are more evident than in previously obtained images of molecular line emission. YSO candidates seem to be spatially associated with L1188a or the intermediate-velocity region rather than L1188b (see Figs.~\ref{Fig:mor}a--\ref{Fig:mor}c). 

Given that the large-scale cloud structures have been investigated in $^{12}$CO (1--0) and $^{13}$CO (1--0) \citep{2017ApJ...835L..14G}, we therefore concentrate on smaller structures in this work. In L1188b, there are at least two long filamentary structures, denoted as F1 and F2 in Fig.~\ref{Fig:mor}f, which are nearly parallel to each other and roughly separated by 0.8--1 pc on a linear scale. F1 appears to be filamentary, while the morphology of F2 displays an oblate ``X''-like shape. Both of them are elongated and measure $\sim$2.7 pc in the northeast-southwest direction. Furthermore, they are roughly perpendicular to the long axis of L1188b, but nearly parallel to the long axis of L1188a. The crest of F1 is obtained from the CO (2--1) integrated intensity map of L1188b (Fig.~\ref{Fig:mor}f) with the discrete persistent structure extractor (DisPerSe\footnote{http://www2.iap.fr/users/sousbie/web/html/indexd41d.html?}) algorithm \citep{2011MNRAS.414..350S,2011MNRAS.414..384S}. Both the persistence and robustness thresholds were set to 3 K~\kms\,to isolate filamentary structures. They were assembled to form a single filament when they form an angle smaller than 70\degree. The resulting crest is shown as the red line in Fig.~\ref{Fig:mor}f.

\begin{figure*}[!htbp]
\centering
\includegraphics[width = 0.95 \textwidth]{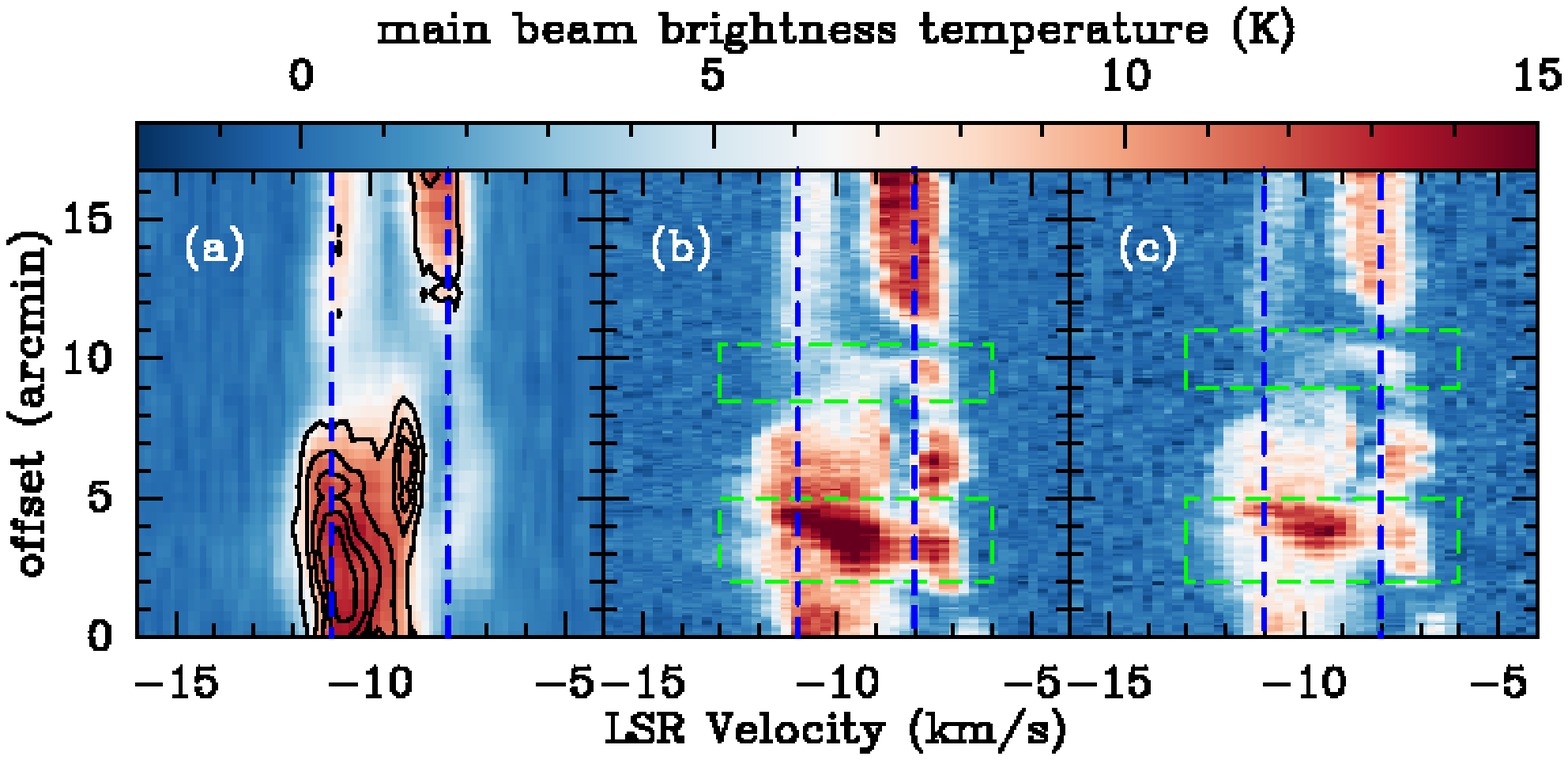}
\includegraphics[width = 0.95 \textwidth]{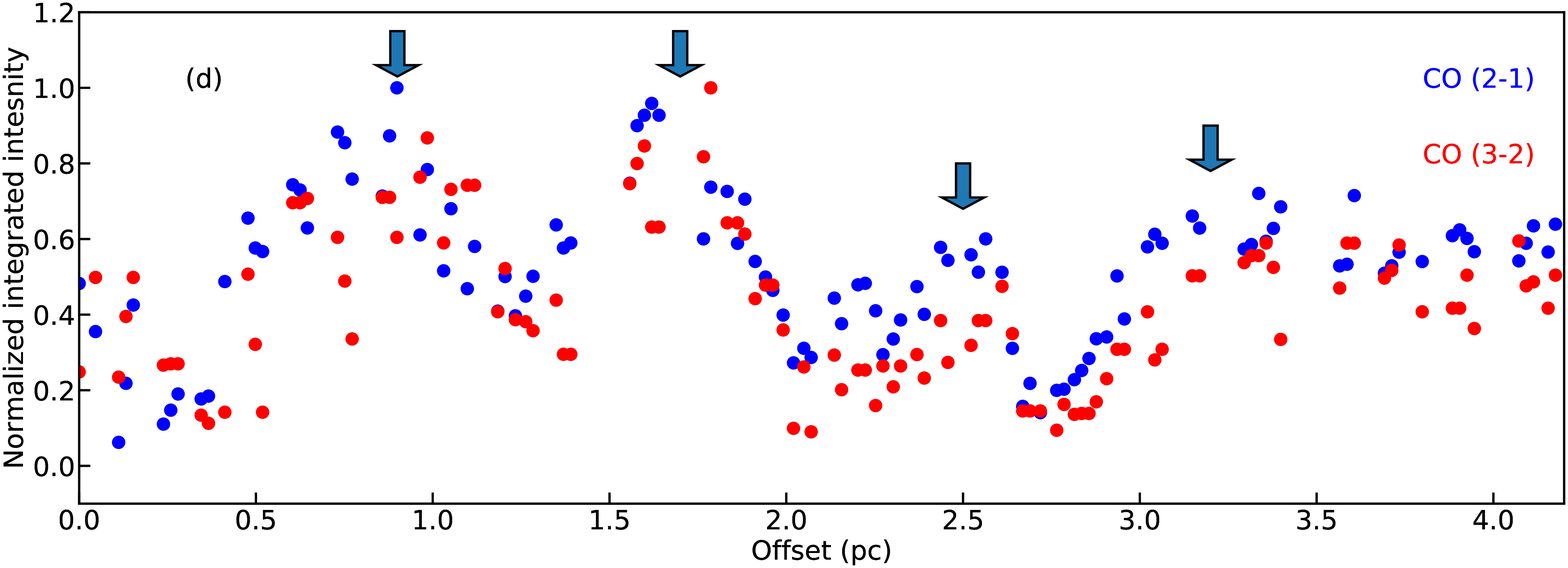}
\caption{{(a) PV diagram of $^{12}$CO (1--0) along the crest of F1 as indicated in Fig.~\ref{Fig:mor}f. The overlaid contours represent the PV diagram of $^{13}$CO (1--0). The contours start at 0.9~K (3$\sigma$) and increase by 0.9 K. The offsets are given with respect to the southwestern end of the PV cut. The systemic velocities of L1188a and L1188b are indicated by two blue vertical dashed lines. (b) Similar to Fig.~\ref{Fig:f1-pv}a, but for $^{12}$CO (2--1). (c) Similar to Fig.~\ref{Fig:f1-pv}a, but for $^{12}$CO (3--2). In Figs.~\ref{Fig:f1-pv}b--\ref{Fig:f1-pv}c, the bridging features are indicated by the green dashed boxes. (d) Normalized integrated intensities of $^{12}$CO (2--1) (blue points) and $^{12}$CO (3--2) (red points) as a function of the offsets in the crest of F1. The offsets are calculated with respect to the southwestern end of F1.}\label{Fig:f1-pv}}
\end{figure*}

Figures~\ref{Fig:f1-pv}a--\ref{Fig:f1-pv}c present the position-velocity (PV) diagram of the CO lines along the crest of of F1. One can clearly distinguish two velocity components with systemic velocities of about $-$11~\kms\,and $-$8~\kms, which well match the systemic velocities of L1188a and L1188b \citep{2017ApJ...835L..14G}. Furthermore, there are two bridging features at offsets of 3.5\arcmin\,and 9.5\arcmin\,that connect these two velocity components (indicated by the two green boxes in Figs.~\ref{Fig:f1-pv}b--\ref{Fig:f1-pv}c), supporting that the two velocity components are associated with each other. Figure~\ref{Fig:f1-pv}d shows the variation of the normalized integrated intensities of CO (2--1) and CO (3--2) along the crest of F1. This variation appears to be periodic with a sine wave-like shape, and one can identify at least four peaks indicated by the arrows in \ref{Fig:f1-pv}d. The four peaks correspond to offsets of 0.9, 1.7, 2.5, and 3.2 pc, respectively, suggesting that the peak-to-peak separations are nearly constant, that is, $\sim$0.8 pc. The maximum-to-minimum ratios are about 2.5-3.0 in this intensity profile.

\begin{figure*}[!htbp]
\centering
\includegraphics[width = 0.95 \textwidth]{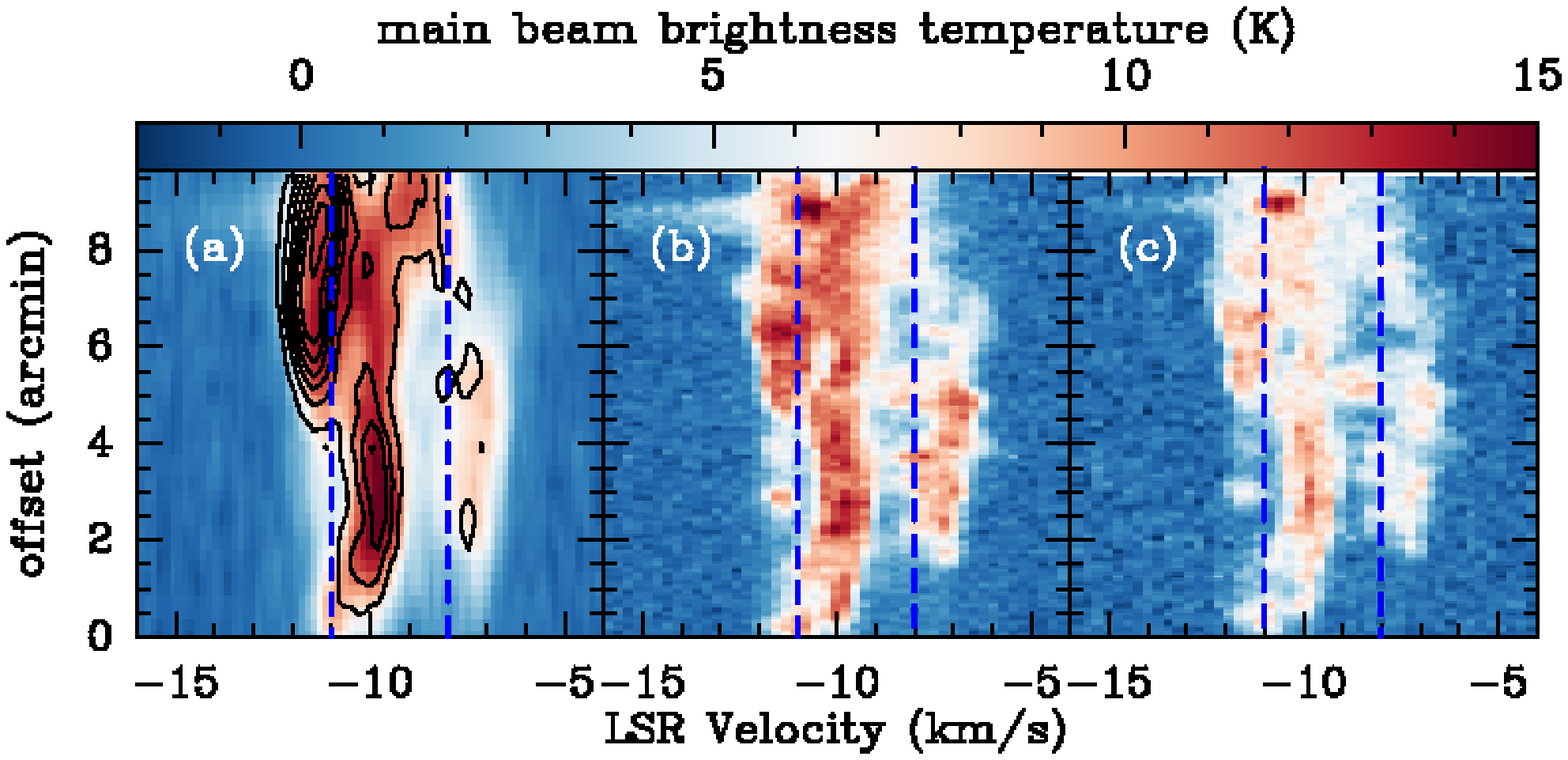}
\caption{{(a) PV diagram of $^{12}$CO (1--0) across F2 indicated by the magenta dashed line in Fig.~\ref{Fig:mor}f. The overlaid contours represent the PV diagram of $^{13}$CO (1--0). The contours start at 0.9~K (3$\sigma$) and increase by 0.9 K. The offsets are given with respect to the southwestern end of the PV cut. The systemic velocities of L1188a and L1188b are indicated by two blue vertical dashed lines. (b) Similar to Fig.~\ref{Fig:f2-pv}a, but for $^{12}$CO (2--1). (c) Similar to Fig.~\ref{Fig:f2-pv}a, but for $^{12}$CO (3--2).}\label{Fig:f2-pv}}
\end{figure*}

Figures~\ref{Fig:f2-pv}a--\ref{Fig:f2-pv}c show the PV diagrams of F2 in the CO lines. One can identify at least three velocity components with LSR velocities of $-$11.5, $-$10, and $-$7~\kms. The systemic velocities of L1188a and L1188b just lie within the gaps of the three velocity components. Furthermore, these three components are connected by bridging features over a linear scale of $\sim$1.1 pc (or 5\arcmin), indicating that they are physically interacting. Consequently, the oblate X-like shape may be due to a superposition of the three components. \citet{2019MNRAS.484.1800S} identify a filamentary YSO distribution indicated in Fig.~\ref{Fig:overview} which is nearly parallel to the long axis of L1188b and perpendicular to the long axis of L1188a, F1, and F2. 

\begin{figure*}[!htbp]
\centering
\includegraphics[width = 0.95 \textwidth]{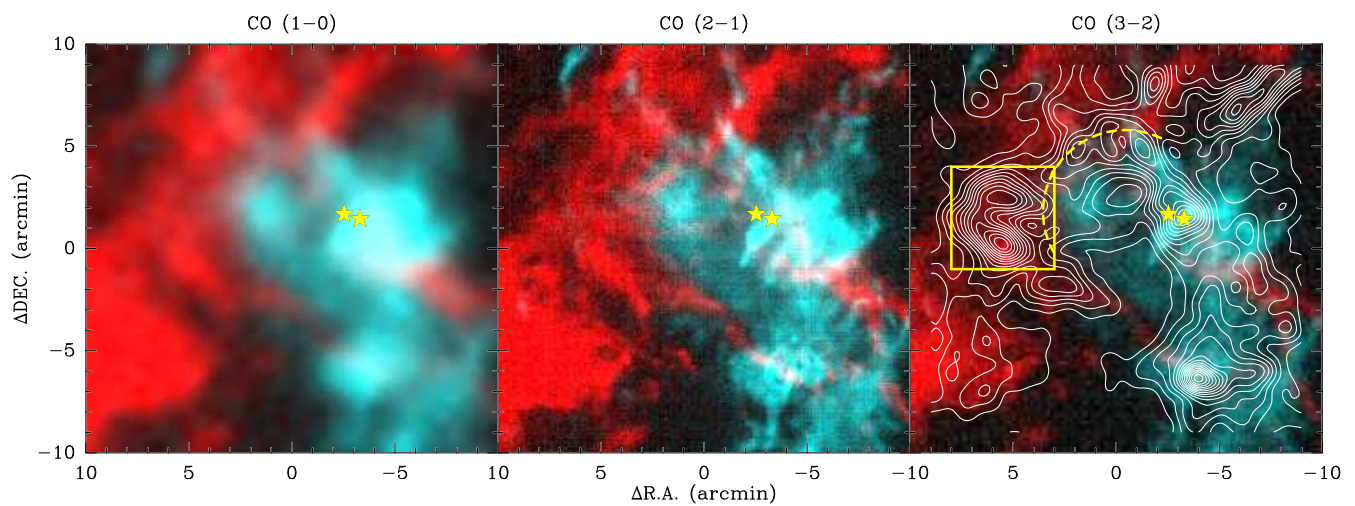}
\caption{{Complementary distribution between L1188a and L1188b observed in $^{12}$CO (1--0) (left), $^{12}$CO (2--1) (middle), and $^{12}$CO (3--2) (right). In all panels, turquoise and red represent the intensity maps integrated from $-$14 to $-$11~\kms\, and from $-$8 to $-$5~\kms, respectively. In the right panel, the overlaid contours are the same as those in Fig.~\ref{Fig:mor}h. The curved region that shows enhanced $^{13}$CO emission is indicated by the yellow box, while the centric concave shape of L1188b is indicated by the yellow dashed arc. In each panel, the two pentagrams represent the two 22 GHz water maser positions discussed in Sect.~\ref{sec.maser}.}\label{Fig:cpt}}
\end{figure*}

\begin{figure}[!htbp]
\centering
\includegraphics[width = 0.45 \textwidth]{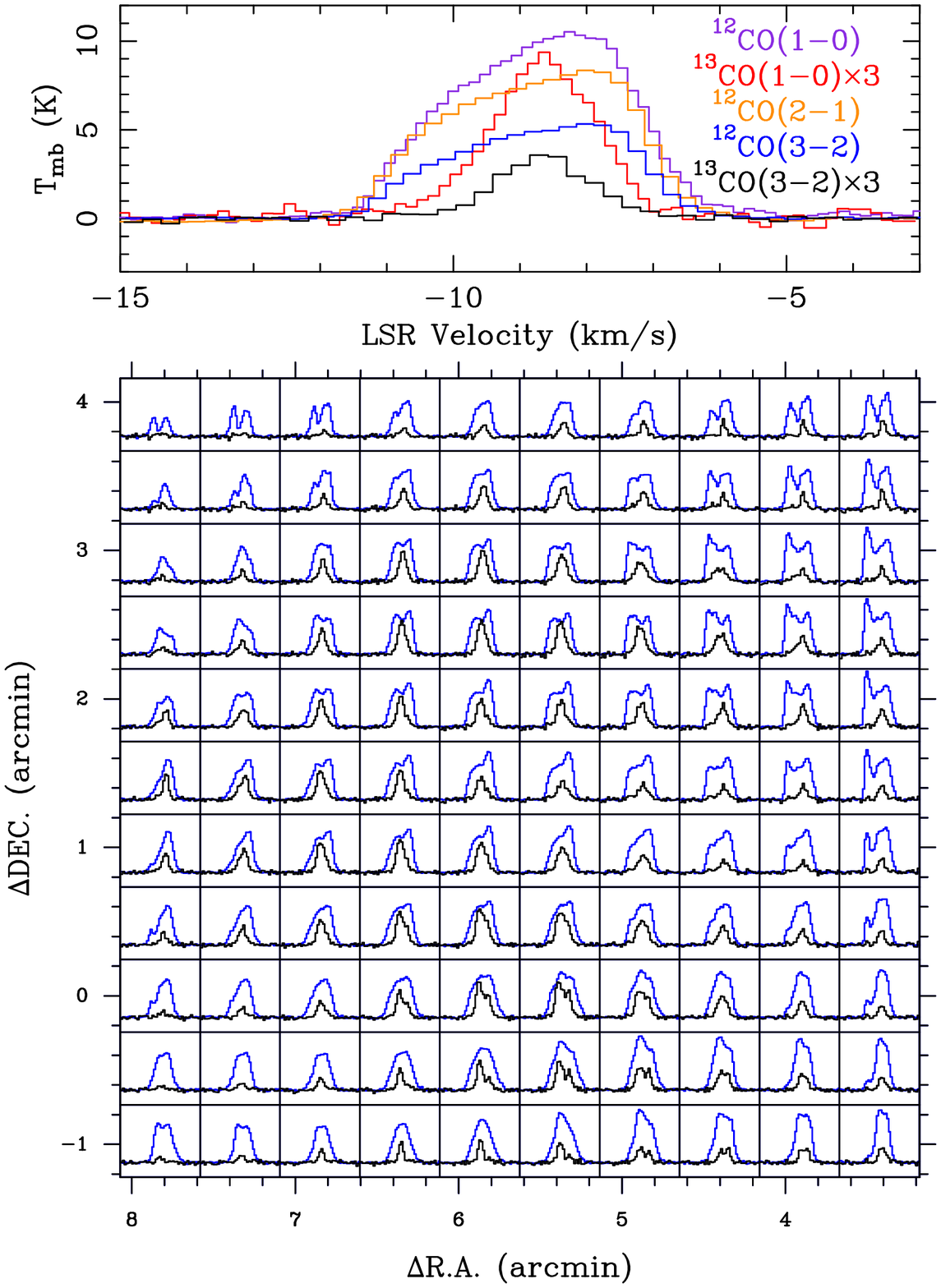}
\caption{{\textit{Top:}$^{12}$CO (1--0), $^{13}$CO (1--0), $^{12}$CO (2--1), $^{12}$CO (3--2), and $^{13}$CO (3--2) spectra averaged over the region indicated by the yellow box in the right panel of Fig.~\ref{Fig:cpt}. \textit{Bottom:} Grid of the $^{12}$CO (3--2) (blue) and $^{13}$CO (3--2) (black) line profiles in the yellow box indicated in Fig.~\ref{Fig:cpt}. These spectra share the same velocity range between $-$15~\kms\,and $-$3~\kms, and intensity range between $-$2~K and 8~K. The $^{13}$CO (3--2) spectra have been scaled by a factor of two. These spectra have been convolved to an angular resolution of 55\arcsec.}\label{Fig:spec}}
\end{figure}

Figure~\ref{Fig:cpt} presents a spatially complementary distribution between L1188a and L1188b. L1188a is sandwiched in L1188b, that is, the northeast end of L1188a matches the centric concave shape of L1188b very well (see the yellow dashed arc in the right panel of Figure~\ref{Fig:cpt}), while F1 matches the southern edge of L1188a. Despite the fact that complementary distributions have also been found in other cloud-cloud collision candidates associated with H{\scriptsize II} regions \citep[e.g.,][]{2017ApJ...835..142T,2018PASJ...70S..51T,2018PASJ...70S..44F,2019ApJ...872...49F}, this is the first time that such a feature is clearly present in a region without massive stars inside. Furthermore, enhanced $^{13}$CO emission is curved in region C3 (see the yellow box in the right panel of Fig.~\ref{Fig:cpt}), which is near the centric concave shape of L1188b. Toward region C3, $^{12}$CO spectra look quite different from $^{13}$CO spectra (see the top panel of Fig.~\ref{Fig:spec}). The $^{12}$CO lines are asymmetric with enhanced redshifted emission with a slope that increases in intensity from lower to higher velocities, while the $^{13}$CO lines appear to be Gaussian. The bottom panel of Fig.~\ref{Fig:spec} shows a grid of the $^{12}$CO (3--2) and $^{13}$CO (3--2) line profiles. Most of the $^{12}$CO (3--2) line profiles display two peaks, while $^{13}$CO (3--2) lines peak toward the dips in the corresponding $^{12}$CO (3--2) lines. Similar line profiles are also observed in $^{12}$CO (1--0), $^{12}$CO (2--1), and $^{13}$CO (1--0). This is indicative of widespread $^{12}$CO self-absorption in this region. 



\subsection{Discovery of a 1 pc-long arc}\label{sec.arc}
\begin{figure*}[!htbp]
\centering
\includegraphics[width = 0.95 \textwidth]{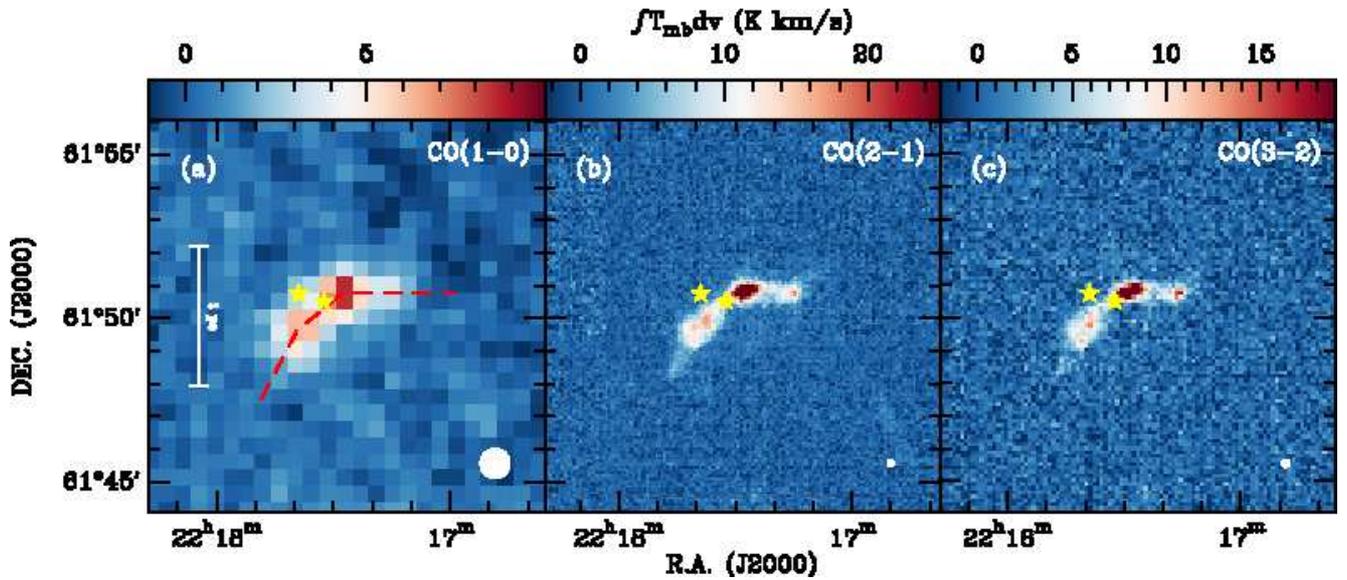}
\caption{{Integrated intensity maps of $^{12}$CO (1--0) (Fig.~\ref{Fig:arc}a), $^{12}$CO (2--1) (Fig.~\ref{Fig:arc}b), and $^{12}$CO (3--2) (Fig.~\ref{Fig:arc}c). All of their integrated velocity ranges extend from $-$20.5 to $-$12.5~\kms. The red dashed line represents the PV cut, and its corresponding PV diagrams are shown in Fig.~\ref{Fig:arc-pv}. In each panel, the two pentagrams represent the two maser positions discussed in Sect.~\ref{sec.maser}. The beam size is shown in the lower right corner of each panel.}\label{Fig:arc}}
\end{figure*}

\begin{figure*}[!htbp]
\centering
\includegraphics[width = 0.95 \textwidth]{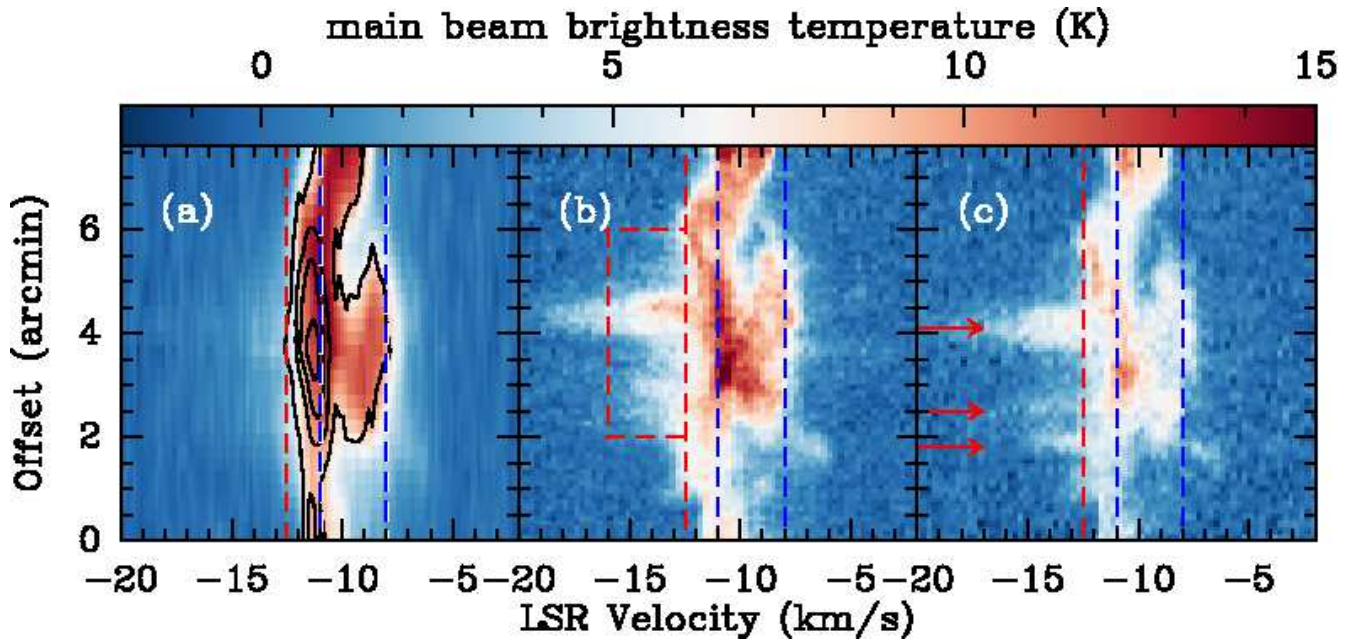}
\caption{{(a) PV diagram of $^{12}$CO (1--0) across the PV cut indicated by the red dashed line in Fig.~\ref{Fig:arc}a. The overlaid contours represent the PV diagram of $^{13}$CO (1--0). The contours start at 0.9~K (3$\sigma$) and increase by 1.8 K. The systemic velocities of L1188a and L1188b are indicated by two blue vertical dashed lines. (b) Similar to Fig.~\ref{Fig:arc-pv}a, but for $^{12}$CO (2--1). The weak low-velocity wing emission is indicated by the red dashed box. (c) Similar to Fig.~\ref{Fig:arc-pv}a, but for $^{12}$CO (3--2). The three arrows represent the high-velocity wing emission discussed in Sect.~\ref{sec.arc}. In all panels, the offsets are given with respect to the southeastern end of the PV cut. The red dashed line is used to highlight the high-velocity wings, which can be recognized on the left hand side of the red dashed line.}\label{Fig:arc-pv}}
\end{figure*}

Figure~\ref{Fig:arc} shows the intensity maps of the three lowest $J$ $^{12}$CO transitions integrated from $-$20.5 to $-$12.5~\kms. This demonstrates that the most blueshifted component exhibits an arc-like filamentary structure which roughly measures $\sim$1 pc in length and $\sim$0.16 pc in width. This results in a high aspect ratio of $\sim$6. This arc is located at the northeastern end of F2, which is nearly perpendicular to the long axis of F2, and extends nearly parallel to the centric concave structure of L1188b. Furthermore, the radius of curvature is estimated to be about 0.6 pc for this arc, which is comparable to the centric concave shape of L1188b. Figure~\ref{Fig:arc-pv} shows the PV diagram along the arc, in which we can identify two velocity components centered at about $-$11~\kms and $-$8~\kms\,, similar to the systemic velocities of L1188a and L1188b, at offsets of 2\arcmin--6\arcmin. The two velocity components are connected to each other by bridging features that extend across the whole arc structure. This indicates that the origin of the 1 pc-long arc could be related to the interaction between L1188a and L1188b. Moreover, weak low-velocity wing emission at velocities of $<-$12.5~\kms\,is also present over the offsets of 2\arcmin--6\arcmin\, and most of the emission can reach $-$16~\kms\,(indicated by the red dashed box in Fig.~\ref{Fig:arc-pv}b). This wing emission is beyond the $^{13}$CO velocity range, and extends $\sim$1 pc in length, indicating that this emission can be caused by large-scale shocks. The high-velocity wing emission is also uncovered, and is indicated by the three red arrows in Fig.~\ref{Fig:arc-pv}c. We find high-velocity wing emission covering a linear scale of $\sim$0.1~pc, which is similar to the typical scale of molecular outflows in low-mass star-forming regions \citep[e.g.,][]{2007prpl.conf..245A}. We also note that these wing emissions are only seen in the blueshifted velocity range, whereas the potential red-shifted emission may mix with the bridging features and the $-$8~\kms\,velocity component, making it undistinguishable in Fig.~\ref{Fig:arc-pv}. At offsets of 6\arcmin--7\arcmin, one can find a velocity gradient of 3~\kms~pc$^{-1}$.

\begin{figure}[!htbp]
\centering
\includegraphics[width = 0.45 \textwidth]{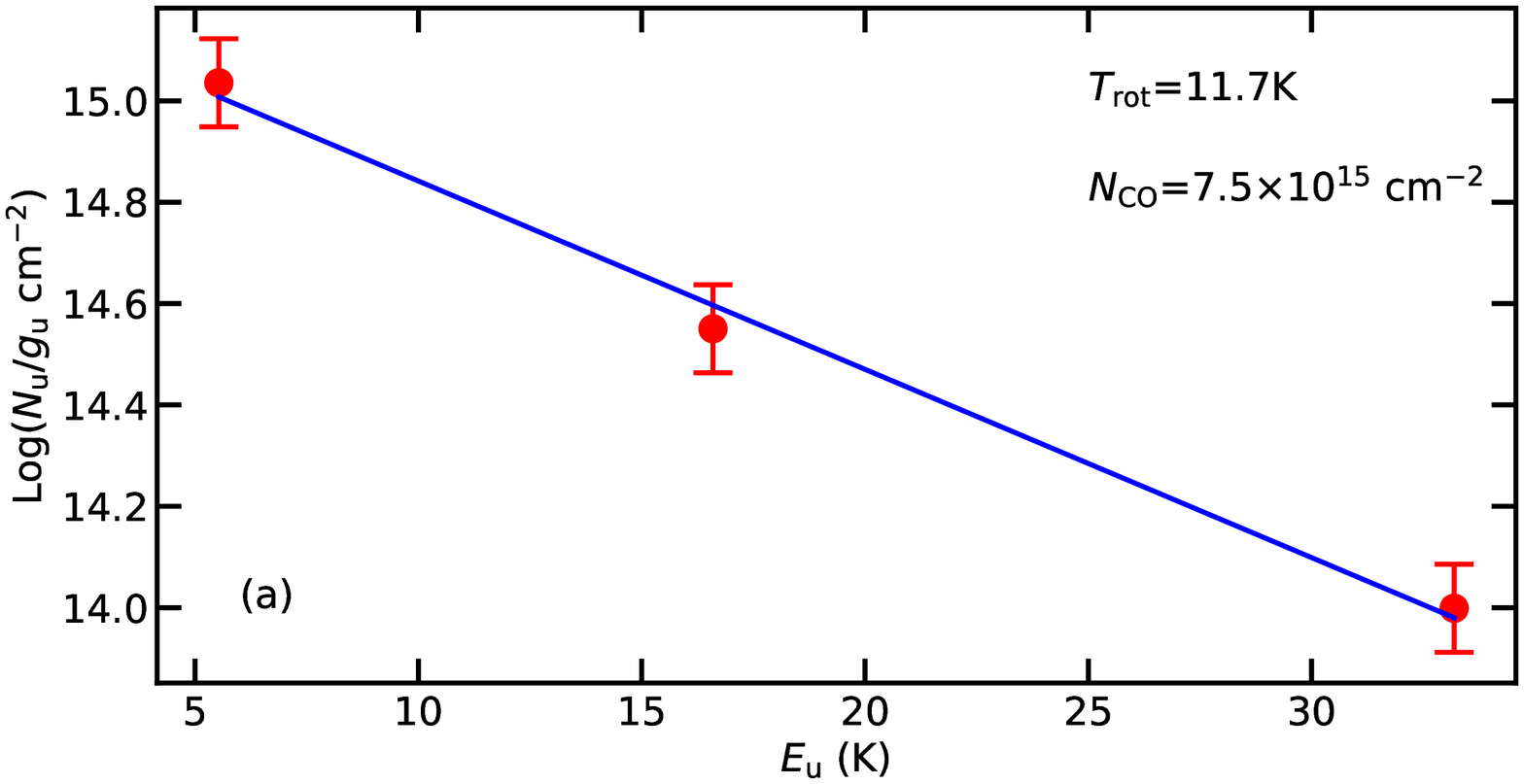}
\includegraphics[width = 0.45 \textwidth]{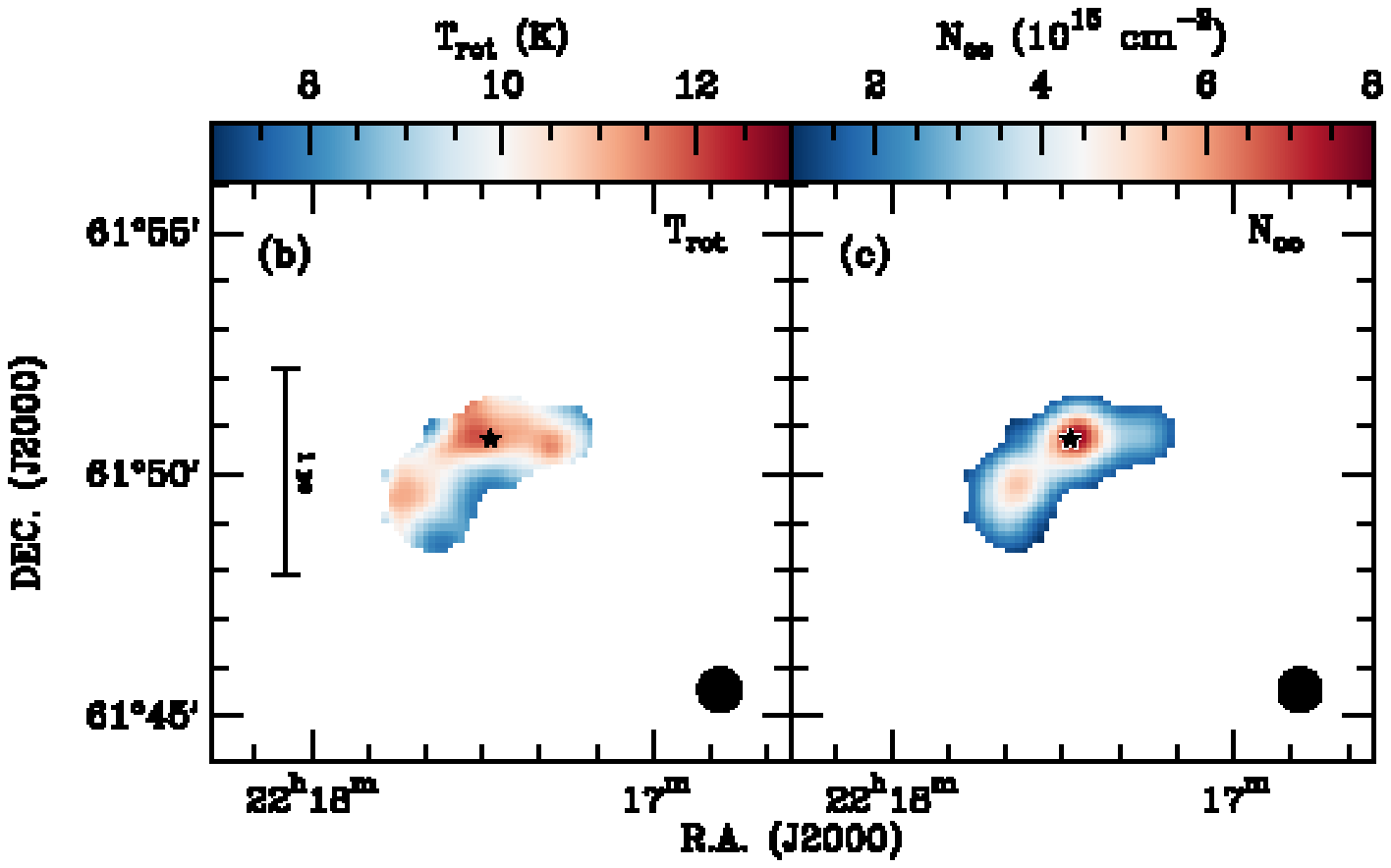}
\caption{{(a) Rotational diagram of the peak position in the arc ($\alpha_{\rm J2000}$=22$^{\rm h}$17$^{\rm m}$28$\rlap{.}^{\rm s}$6, $\delta_{\rm J2000}$=61\degree50\arcmin44$\rlap{.}^{\prime\prime}$8) indicated by the pentagram in Figs.~\ref{Fig:arc-ext}b--\ref{Fig:arc-ext}c. (b) Distribution of rotational temperatures in the arc structure. (c) Distribution of derived $^{12}$CO column densities in the arc structure. The beam size is shown in the lower right corner of Figs.~\ref{Fig:arc-ext}b and \ref{Fig:arc-ext}c. }\label{Fig:arc-ext}}
\end{figure}

Assuming that the CO wing emission integrated from $-$20.5 to $-$12.5~\kms\,is optically thin and that it follows local thermodynamic equilibrium (LTE), we are able to derive $^{12}$CO column densities and rotational temperatures in the arc structure with the rotational diagram. Prior to the rotational diagram analysis, we convolved $^{12}$CO (2--1) and $^{12}$CO (3--2) data to the angular resolution of 55\arcsec, and all data were then linearly interpolated to the same grid. We used the following formula \citep[e.g.,][]{2009ARA&A..47..427H,2015A&A...574A..56G}:
\begin{equation}\label{f.rot}
{\rm ln}(\frac{3kW}{8\pi^{3}\nu\mu^{2}S})~=~{\rm ln}(\frac{N_{\rm tot}}{Q(T_{\rm rot})})-\frac{E_{\rm u}}{kT_{\rm rot}},
\end{equation}
where $k$ is the Boltzmann constant, $W$ is the $^{12}$CO integrated intensity, $\nu$ is the rest frequency, $\mu$ is the permanent dipole moment, $S$ is the transition's intrinsic strength, $N_{\rm tot}$ is the total column density, $T_{\rm rot}$ is the rotational temperature, $Q$ is the partition function, and $E_{\rm u}$ is the upper level energy of the transition. The values of $Q$ and $\mu$ are taken from the CDMS\footnote{https://zeus.ph1.uni-koeln.de/cdms/catalog} \citep{2005JMoSt.742..215M}. A linear fit was performed toward every pixel to obtain their $^{12}$CO rotational temperatures and column densities, and an example toward the peak in the arc structure is shown in Fig.~\ref{Fig:arc-ext}a. In the following analysis, we only account for pixels with at least 5$\sigma$ in all three $^{12}$CO transitions. The resulting $^{12}$CO rotational temperature and column density maps are shown in Figs.~\ref{Fig:arc-ext}b--\ref{Fig:arc-ext}c. The derived rotational temperatures range from 8 K to 12 K, while $^{12}$CO column densities are within the range of (1.1--7.6)$\times 10^{15}$~cm$^{-2}$. Assuming a typical $^{12}$CO abundance of 8$\times 10^{-5}$ \citep{1982ApJ...262..590F} and the mean molecular weight of 2.8 per hydrogen molecule \citep[e.g.,][]{2008A&A...487..993K}, the total gas mass of the arc structure is about 0.34~M$_{\odot}$ above the 5$\sigma$ threshold. We also note that this arc is not detected in $^{13}$CO lines. The 3$\sigma$ upper limit of the $^{13}$CO (1--0) integrated intensity is 1.2~K~\kms, corresponding to a $^{13}$CO column density of 6$\times 10^{14}$ cm$^{-2}$ at an assumed excitation temperature of 10 K under LTE conditions. 




\subsection{Water maser emission in L1188}\label{sec.maser}

\begin{figure*}[!htbp]
\centering
\includegraphics[width = 0.95 \textwidth]{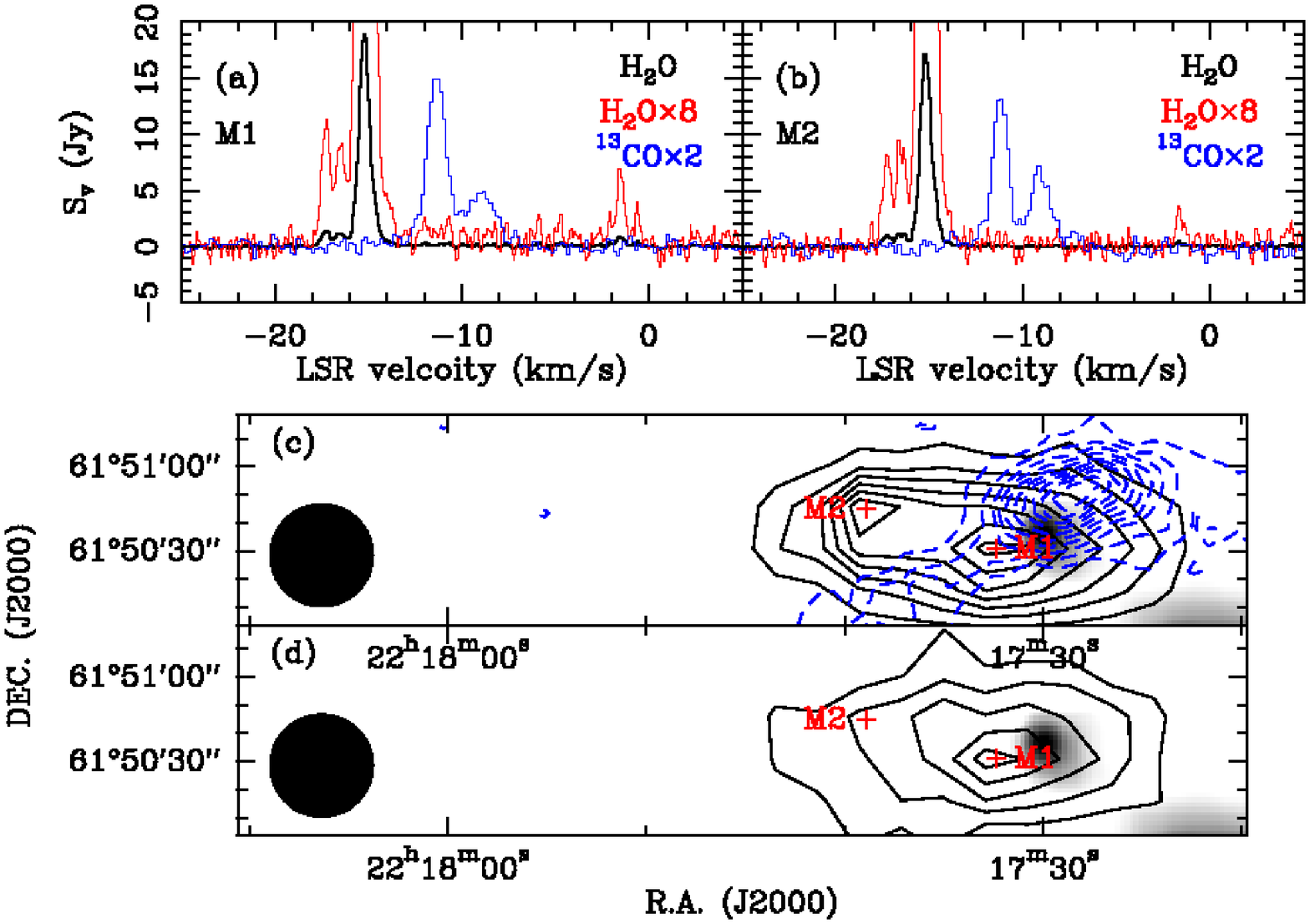}
\caption{{(a) Observed 22 GHz H$_{2}$O maser (black) and $^{13}$CO (1--0) (blue) spectra toward the position M1. (b) Similar to Fig.~\ref{Fig:maser}a, but for the position M2. In Figs.~\ref{Fig:maser}a--\ref{Fig:maser}b, the red lines represent the 22 GHz H$_{2}$O maser lines, which are scaled by a factor of eight in order to better visualize weak velocity components. (c) 22 GHz H$_{2}$O maser integrated intensity contours (black) overlaid on the WISE 22~$\mu$m image. The integrated velocity range is from $-$18 to $-$14~\kms. The black contours start at 2~Jy~\kms\, and increase by 2~Jy~\kms. The blue dashed contours represent the $^{12}$CO (3--2) integrated intensities which start at 2.4~K~\kms\,(3$\sigma$) and increase by 2.4~K~\kms. (d) 22 GHz H$_{2}$O maser integrated intensity contours overlaid on the WISE 22~$\mu$m image. The integrated velocity range is from $-$2 to $-$0~\kms. The contours start from 0.1~Jy~\kms\, and increase by 0.1~Jy~\kms. In Figs.~\ref{Fig:maser}c--\ref{Fig:maser}d, the beam size is shown in the lower left corner of each panel.}\label{Fig:maser}}
\end{figure*}

Theoretical models have predicted that water masers can be created by cloud-cloud collisions \citep{1986ApJ...305..467T}, but direct observational evidence is still rare. We searched for 22 GHz water maser emission in a 360\arcsec$\times$75\arcsec\, area of region C1 close to the 1 pc-long shocked arc (see Figs.~\ref{Fig:arc}--\ref{Fig:arc-ext}), and detected water maser emission in two positions. This is the first maser detection in L1188. Their spectra and spatial distributions are shown in Figs.~\ref{Fig:maser}a--Figs.~\ref{Fig:maser}d, and the two peak positions, M1 and M2, were mannually determined by the naked eyes. The observed maser properties are tabulated in Table~\ref{Tab:maser}.

The 22 GHz water maser spectral profiles are different from (quasi-)thermal line profiles of other molecules. The former ones display at least four peaks, whereas the latter ones only show two peaks at $-$11~\kms\,and $-$9~\kms\,(see Figs.~\ref{Fig:maser}a--\ref{Fig:maser}b). The peak velocities of the maser spectra are $\gtrsim$5\kms\,off the $^{13}$CO (1--0) peak velocities. These peaks in the maser spectra can be divided into two groups separated by a velocity difference of $>$11~\kms, that is, a blueshifted group at a velocity range between $-$18~\kms\,and $-$13~\kms, and a redshifted group at a velocity range between $-$2~\kms\,and 0~\kms. The strongest maser component has an LSR velocity of about $-$15.2~\kms, which is blueshifted with respect to the systemic velocities of L1188a and L1188b that are traced by $^{13}$CO. Its corresponding main beam brightness temperature is higher than the peak intensities of the three $^{12}$CO lines. A comparison of the two linearly polarized components of these maser features shows no significant differences in line shape and flux density. Their isotropic H$_{2}$O luminosities reach $\sim$2$\times 10^{-7}$~L$_{\odot}$, which is nearly an order of magnitude higher than those in nearby low-mass star-forming regions \citep{2003ApJS..144...71F}, but lower than those in most high-mass star-forming regions \citep{2011MNRAS.418.1689U}.

Figures~\ref{Fig:maser}c--\ref{Fig:maser}d show that the 22 GHz water maser emission mainly arises from two positions, M1 and M2, which are indicated by the two red crosses. They are separated by 48\arcsec, corresponding to $\sim$0.2 pc. Near M1, we find the Class I YSO WISEJ221729.87+615039.8 (details are described in Appendix~\ref{sec.app}), which is seen in emission at 22~$\mu$m (see Figures~\ref{Fig:maser}c--\ref{Fig:maser}d). The separations between this YSO and the two maser positions are about 20\arcsec\,(or 0.08 pc) and 60\arcsec\,(or 0.23 pc), respectively. The offsets and the spectral profiles are similar to those found in other regions where masers are excited by YSO outflows \citep{2001ApJ...559L.143F,2003ApJS..144...71F}. This indicates that the detected masers could be excited by YSO outflows rather than cloud-cloud collisions. However, we cannot resolve outflows from this YSO in its northeastern direction where masers are located. This is likely due to insufficient sensitivities. Since previous distance measurements of L1188 ($\sim$800 pc, see Sect.~\ref{sec:info}) still have large uncertainties \citep{2017ApJ...835L..14G,2019MNRAS.484.1800S}, future trigonometric parallax measurements of the newly detected 22 GHz water masers with Very Long Baseline Interferometry (VLBI) measurements may provide distance estimates of a much higher quality.

\begin{table*}[!hbt]
\caption{Observed properties of the two detected 22 GHz water maser positions.}\label{Tab:maser}
\normalsize
\centering
\begin{tabular}{cccccccc}
\hline \hline
maser          & (R.A. (J2000))                & DEC. (J2000)                &$\varv_{\rm lsr}$-range       & $\varv_{\rm p}$     & $S_{\nu}$   & $\int S_{\nu}{\rm d}\varv$ & $L_{\rm H2O}$  \\
               & ($^{\rm h}$~$^{\rm m}$~$^{\rm s}$) & (\degree\,\arcmin\,\arcsec) &  (\kms)                    & (\kms)            & (Jy)       &  (Jy~\kms)                & ($L_{\odot}$)    \\
(1)            & (2)                           & (3)                         & (4)                        & (5)               &  (6)       &  (7)                      & (8)   \\
\hline
M1             & 22:17:32.4                    & 61:50:31                    & [$-$18.0,$-$16.9]          & $-$17.2           & 1.4$\pm$0.1 & 0.75$\pm$0.03            &  1.1$\times 10^{-8}$   \\
               & 22:17:32.4                    & 61:50:31                    & [$-$16.9,$-$16.0]          & $-$16.5           & 1.1$\pm$0.1 & 0.80$\pm$0.02            &  1.2$\times 10^{-8}$   \\
               & 22:17:32.4                    & 61:50:31                    & [$-$16.0,$-$13.4]          & $-$15.2           & 18.9$\pm$0.1& 14.22$\pm$0.04           &  2.1$\times 10^{-7}$   \\
               & 22:17:32.4                    & 61:50:31                    & [$-$2.0,$-$1.0]            & $-$1.5            & 0.9$\pm$0.1 & 0.28$\pm$0.02            &  4.2$\times 10^{-9}$   \\
               & 22:17:32.4                    & 61:50:31                    & [$-$0.8,$-$0.3]            & $-$0.6            & 0.5$\pm$0.1 & 0.01$\pm$0.02            &  1.5$\times 10^{-10}$   \\
\hline
M2             & 22:17:38.9                    & 61:50:45                    & [$-$17.8,$-$16.9]          & $-$17.2           & 1.0$\pm$0.1 & 0.53$\pm$0.02            &  7.9$\times 10^{-9}$   \\
               & 22:17:38.9                    & 61:50:45                    & [$-$16.9,$-$16.0]          & $-$16.6           & 1.1$\pm$0.1 & 0.75$\pm$0.02            &  1.1$\times 10^{-8}$   \\
               & 22:17:38.9                    & 61:50:45                    & [$-$16.0,$-$13.4]          & $-$15.2           & 17.2$\pm$0.1& 13.46$\pm$0.04           &  2.0$\times 10^{-7}$  \\
               & 22:17:38.9                    & 61:50:45                    & [$-$2.0,$-$1.0]            & $-$1.7            & 0.4$\pm$0.1 & 0.19$\pm$0.02            &  2.8$\times 10^{-9}$  \\
\hline
\end{tabular}
\tablefoot{(1) The maser position. (2) Right ascension. (3) Declination. (4) The corresponding velocity range. (5) The peak velocity. (6) The peak flux density. (7) The flux density integrated over the corresponding velocity range. (8) The 22 GHz water maser luminosity.}
\normalsize
\end{table*}

\subsection{Non-detection of SiO (2--1) and CH$_{3}$OH ($8_{0}$--$7_{1}$ A$^{+}$) emission}
Since SiO (2--1) is a commonly used shock tracer \citep[e.g.,][]{1992A&A...254..315M}, its emission may trace the shock created by the cloud-cloud collision. The CH$_{3}$OH ($8_{0}$--$7_{1}$ A$^{+}$) line is a well-known Class I methanol maser transition that is widespread in the Milky Way \citep[e.g.,][]{2017ApJS..231...20Y} and frequently found to be associated with molecular outflows from YSOs. However, it has also been found to occur in supernova remnant molecular cloud interactions \citep{2011MmSAI..82..703F}, and it is thought to be able to form via cloud-cloud collisions \citep[e.g.,][]{1996IAUS..178..163M,2016A&A...592A..31L}. We therefore searched for emission in the SiO and CH$_{3}$OH lines in the region in which emission from other molecules is detected, but we were only able to establish an upper limit of 0.54~K~\kms\,($3\sigma$) for the intensity of SiO (2--1) integrated from $-$14~\kms to $-5$~\kms. The $3\sigma$ upper limit for the SiO column density is $\sim$1.2$\times 10^{12}$~cm$^{-2}$ by assuming LTE and an excitation temperature of 10~K. The upper limit of the main beam brightness temperature for CH$_{3}$OH ($8_{0}$--$7_{1}$ A$^{+}$) is 0.24~K ($3\sigma$), corresponding to a peak flux density of 9.8 Jy.

\section{Discussion}
\subsection{Line ratios and physical properties}\label{sec.phy}
\begin{figure*}[!htbp]
\centering
\includegraphics[width = 0.95 \textwidth]{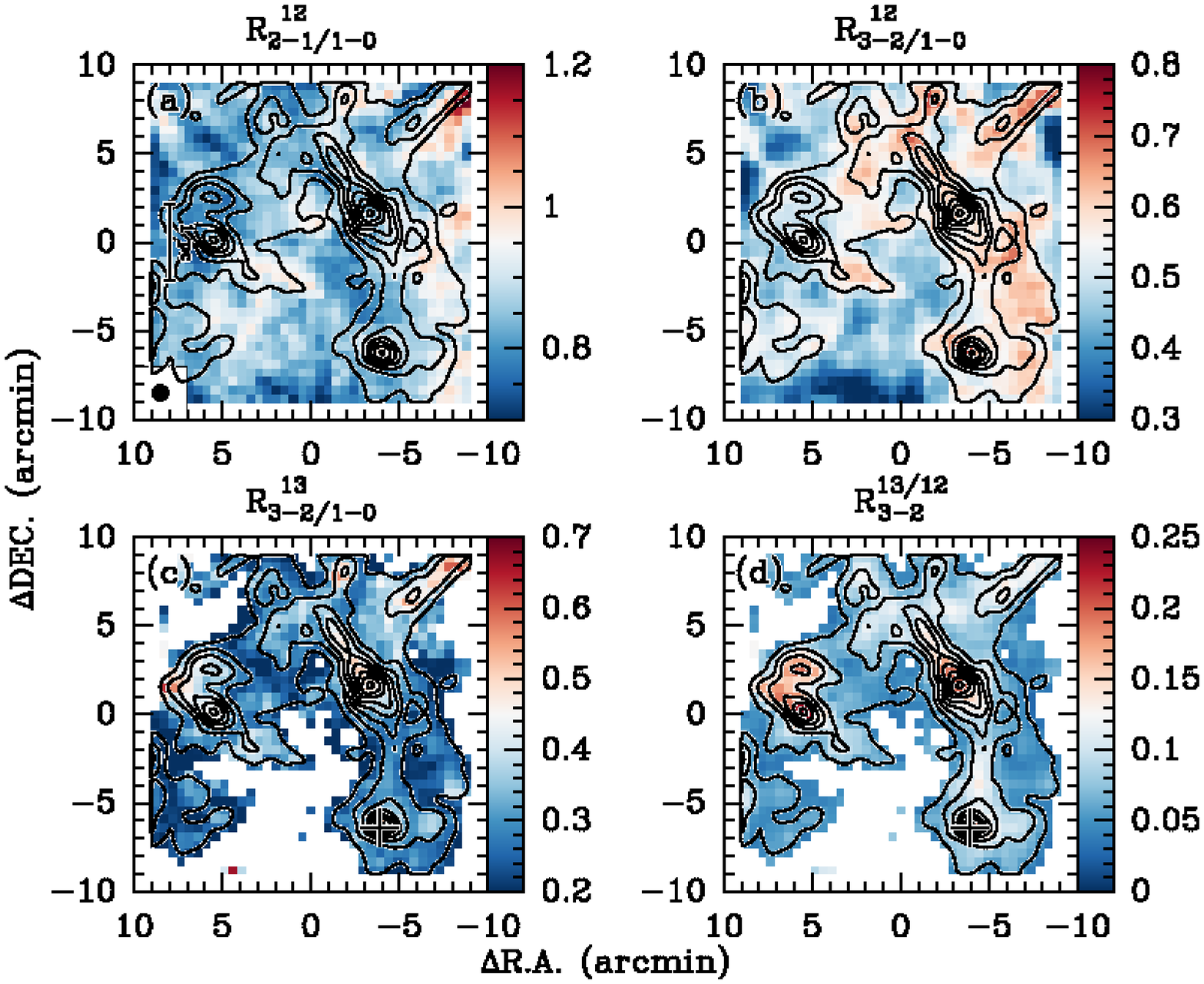}
\caption{{Line ratio $R^{12}_{2-1/1-0}$ (Fig.~\ref{Fig:ratio}a), $R^{12}_{3-2/1-0}$ (Fig.~\ref{Fig:ratio}b), $R^{13}_{3-2/1-0}$ (Fig.~\ref{Fig:ratio}c), and $R^{13/12}_{3-2}$ (Fig.~\ref{Fig:ratio}d) distribution of the observed region. The overlaid contours are $^{13}$CO (3--2) integrated intensities, which start from 0.9 K~\kms\,(6$\sigma$) and increase by 0.9~K~\kms. The beam sizes are shown in the lower left corner of Figs.~\ref{Fig:ratio}a. In all panels, the (0, 0) offset corresponds to $\alpha_{\rm J2000}$=22$^{\rm h}$18$^{\rm m}$00$\rlap{.}^{\rm s}$44, $\delta_{\rm J2000}$=61\degree49\arcmin03$\rlap{.}^{\prime\prime}$7. In Figs.~\ref{Fig:ratio}c--\ref{Fig:ratio}d, the cross represents the peak position of region C2 that is analyzed in Fig.~\ref{Fig:chi}.}\label{Fig:ratio}}
\end{figure*}

CO line ratios are widely used to infer physical properties of molecular gas in the Milky Way and external galaxies \citep[e.g.,][]{2010ApJ...724.1336M,2011ApJ...738...46T,2014A&A...568A.122Z,2015ApJS..216...18N,2017ApJ...835..142T,2018MNRAS.475.1508P,2019A&A...631A.110Z}. In order to derive line ratios in L1188, we convolved the data of all of the observed $^{12}$CO and $^{13}$CO transitions to the same angular resolution and then interpolated them into the same grid as $^{13}$CO (1--0). Here, we use $R^{12}_{2-1/1-0}$, $R^{12}_{3-2/1-0}$, $R^{13}_{3-2/1-0}$, and $R^{13/12}_{3-2}$ to represent the line ratios $^{12}$CO (2--1)/$^{12}$CO (1--0), $^{12}$CO (3--2)/$^{12}$CO (1--0), $^{13}$CO (3--2)/$^{13}$CO (1--0), and $^{13}$CO (3--2)/$^{12}$CO (3--2), respectively. The line ratio maps were derived from the intensity maps integrated from $-$14 to $-$5~\kms. Here, we only took the pixels with at least 5$\sigma$ detections into account. The line ratio maps of the observed region are shown in Fig.~\ref{Fig:ratio}. It is important to note that $R^{12}_{2-1/1-0}$ varies from 0.7 to 1.2 with a median value of 0.9, suggesting a small variation in $R^{12}_{2-1/1-0}$. This is likely because both $^{12}$CO (1--0) and $^{12}$CO (2--1) have very high opacities and nearly the same excitation temperature. Additionally, $R^{12}_{3-2/1-0}$ lies within a range of 0.2--0.7 with a median value of 0.5, generally lower than $R^{12}_{2-1/1-0}$. This can be due to $^{12}$CO (3--2)'s relatively lower opacity and/or its subthermal state. Furthermore, $R^{13}_{3-2/1-0}$ is found to range from 0.1 to 0.7. Since both $^{13}$CO (3--2) and $^{13}$CO (1--0) are likely optically thin, $R^{13}_{3-2/1-0}$ allows us to determine the excitation temperature in case of LTE. The observed ratios suggest an excitation temperature range of 6--10 K. However, the excitation temperature should be lower than the kinetic temperature since $^{13}$CO (3--2) becomes subthermal (see discussions below). When the variation of $^{12}$C/$^{13}$C isotopic ratios can be neglected, $R^{13/12}_{3-2}$ is sensitive to CO column densities and number densities. We find that the $R^{13/12}_{3-2}$ distribution is quite similar to that of the $^{13}$CO emission; $R^{13/12}_{3-2}$ varies from 0.03 to 0.23. The highest $R^{13/12}_{3-2}$ corresponds to region C3, which shows enhanced $^{13}$CO (3--2) emission and $^{12}$CO (3--2) self-absorption (Figs.~\ref{Fig:cpt}--\ref{Fig:spec}).

In order to derive physical properties, we employed a non-LTE analysis using the RADEX\footnote{https://personal.sron.nl/~vdtak/radex/index.shtml} code \citep{2007A&A...468..627V}. The Einstein A coefficients are from the Leiden Atomic and Molecular Database \citep[LAMDA\footnote{https://home.strw.leidenuniv.nl/~moldata/},][]{2005A&A...432..369S}, while collision rates are taken from \citet{2010ApJ...718.1062Y}. An expanding spherical geometry, also known as large velocity gradient approximation \citep{1960mes..book.....S}, has been assumed for the calculations, so the escape probability ($\beta$) is $\beta=\frac{1-{\rm exp}(-\tau)}{\tau}$, where $\tau$ is the optical depth. The ortho-to-para ratio is fixed to three for H$_{2}$. A total of 41 $^{12}$CO energy levels and 41 $^{13}$CO energy levels are included in the calculations. Since $^{12}$CO (1--0) and $^{12}$CO (2--1) tend to have very high opacities indicated by $R^{12}_{2-1/1-0}\sim$1, we only used $R^{13}_{3-2/1-0}$ and $R^{13/12}_{3-2}$ to model physical properties by assuming that the three transitions can trace almost the same portion of molecular gas, following the prescription of previous studies \citep[e.g.,][]{2015ApJS..216...18N}. In order to fit the two ratios simultaneously, we assumed the isotopic ratio $^{12}$C/$^{13}$C to be 70 \citep[e.g.,][]{2015A&A...581A..48G,2016ApJ...824..136L}, the $^{12}$CO abundance to be 8$\times 10^{-5}$ \citep{1982ApJ...262..590F}, and the velocity gradient ${\rm d}\varv/{\rm d}r$ to be 3~\kms~pc$^{-1}$, which is based on the typical $^{13}$CO line width and size of the cloud. During modeling, we considered a regular grid with kinetic temperature ranging from 5 to 100 K with a step of 1 K and the H$_{2}$ number density log[$n$(H$_{2}$) cm$^{-3}$] varying from 2 to 6 with a step of 0.1. In order to obtain physical conditions for the observed region, we make use of the $\chi^{2}$ analysis to evaluate different models toward every pixel. We note that $\chi^{2}$ is defined as
\begin{equation}\label{f.chi}
\chi^{2} = \Sigma_{\rm i} \frac{(R_{\rm obs(i)}- R_{\rm model(i)})^{2}}{\sigma_{\rm obs(i)}^{2}}\;,
\end{equation}
where $R_{\rm obs(i)}$ and $R_{\rm model(i)}$ represent the observed and modeled line ratios. Also, $\sigma_{\rm obs(i)}$ represents the 1$\sigma$ error in $R_{\rm obs(i)}$, which includes the flux calibration uncertainties and is calculated with the error propagation formula. The best fit is derived by minimizing $\chi^{2}$. The 1$\sigma$ confidence range can be derived by $\chi^{2} <$2.3 with two free parameters.  

\begin{figure}[!htbp]
\centering
\includegraphics[width = 0.45 \textwidth]{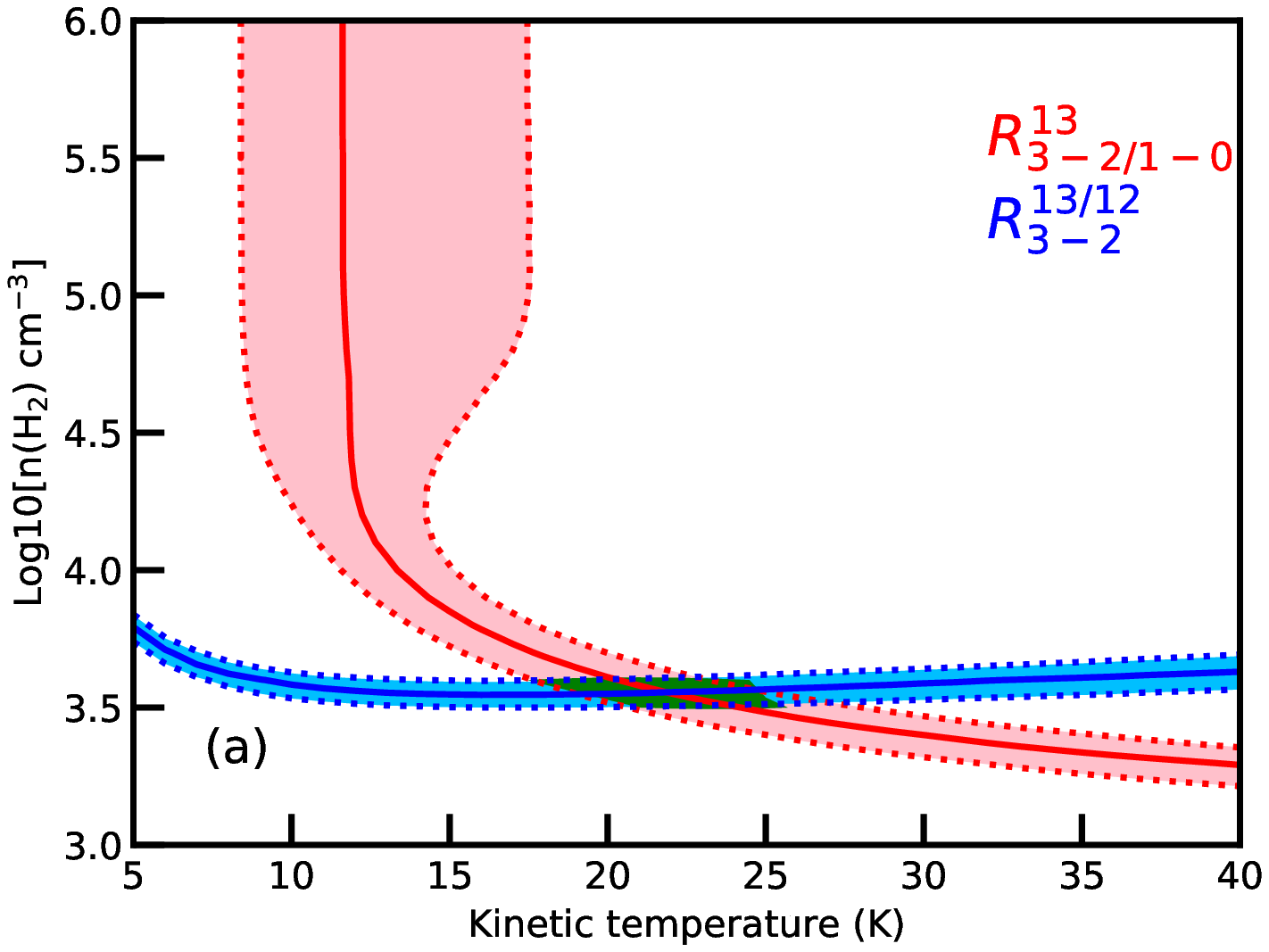}
\includegraphics[width = 0.45 \textwidth]{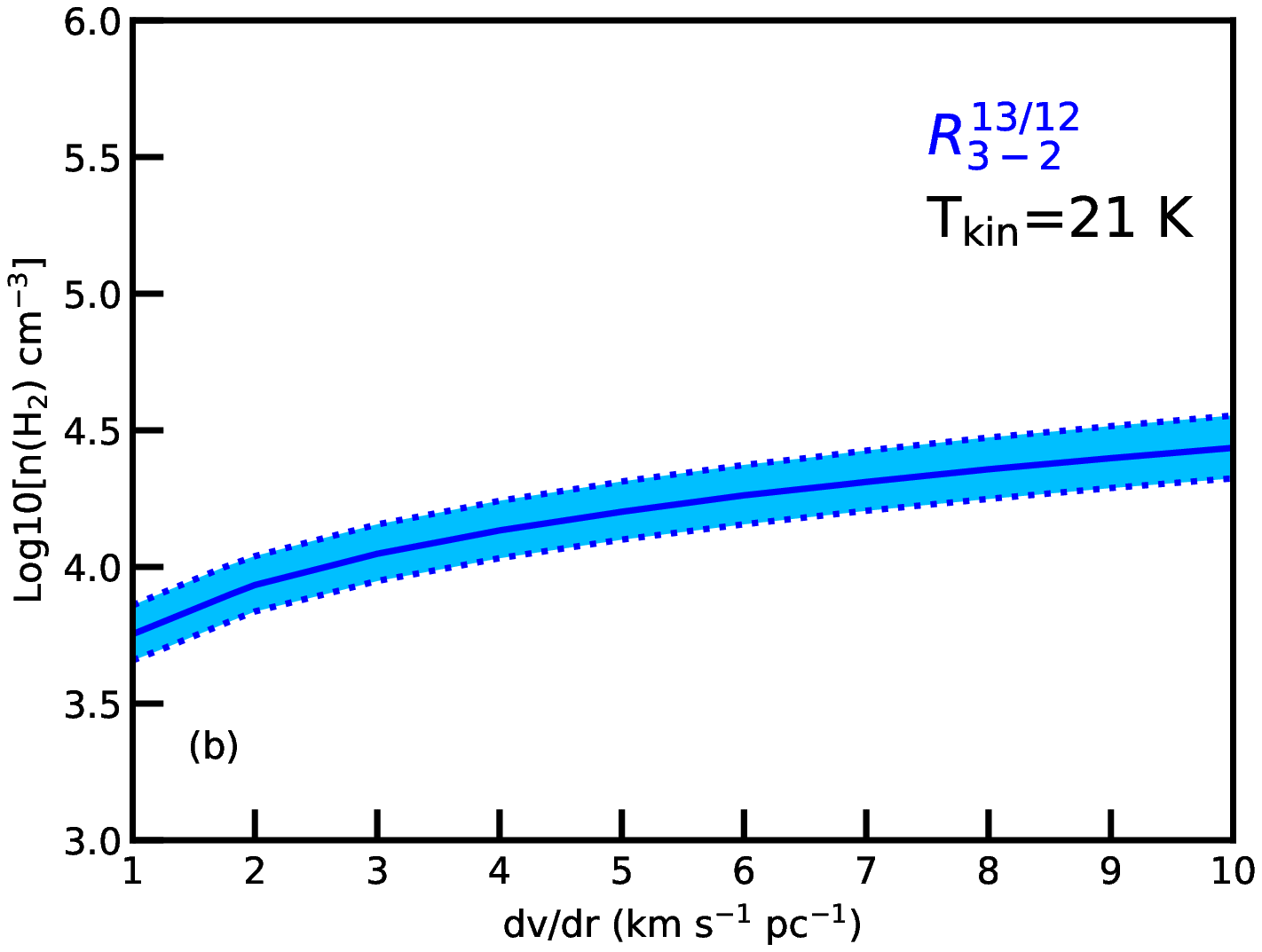}
\caption{{RADEX calculations on modeling the line ratios toward the peak position in region C2 ($\alpha_{\rm J2000}$=22$^{\rm h}$17$^{\rm m}$27$\rlap{.}^{\rm s}$4, $\delta_{\rm J2000}$=61\degree42\arcmin42$^{\prime\prime}$) indicated by the cross in Figs.~\ref{Fig:ratio}c--\ref{Fig:ratio}d. (a) The modeled line ratios, $R^{13}_{3-2/1-0}$ (red) and $R^{13/12}_{3-2}$ (blue), as a function of kinetic temperature and H$_{2}$ number density. The solid lines and the dotted lines represent the observed line ratios and their 1$\sigma$ uncertainties for $R^{13}_{3-2/1-0}$ and $R^{13/12}_{3-2}$. The shadowed region in the intersection represents $\chi^{2}<$2.3, corresponding to the 1$\sigma$ range. (b) The modeled $R^{13/12}_{3-2}$ as a function of H$_{2}$ number density at a given kinetic temperature of 21~K.}\label{Fig:chi}}
\end{figure}

Figure~\ref{Fig:chi}a presents an example of the modeling results toward the peak position of region C2 indicated by the cross in Figs.~\ref{Fig:ratio}c--\ref{Fig:ratio}d. We can see that the kinetic temperature and H$_{2}$ number density are well constrained to be 18--25 K and 10$^{3.5-3.7}$ cm$^{-3}$, respectively, within the 1$\sigma$ confidence range, while the best fit is 21~K for the kinetic temperature and 10$^{3.6}$ cm$^{-3}$ for the H$_{2}$ number density. Our results also suggest that $R^{13/12}_{3-2}$ can be regarded as a reliable tracer to derive H$_{2}$ number densities because of its weak dependence on kinetic temperature. However, the assumed ${\rm d}\varv/{\rm d}r$ may have a large uncertainty. Since $^{13}$CO (3--2) and $^{13}$CO (1--0) are likely optically thin in the observed region, the ${\rm d}\varv/{\rm d}r$ values do not affect the modeled $R^{13}_{3-2/1-0}$ ratio. In order to quantify the impact of the variation of ${\rm d}\varv/{\rm d}r$ on $R^{13/12}_{3-2}$, we modeled $R^{13/12}_{3-2}$ with a ${\rm d}\varv/{\rm d}r$ range of 1--10~\kms~pc$^{-1}$ and a fixed kinetic temperature of 21~K. The result is shown in Fig.~\ref{Fig:chi}b. This suggests that the derived H$_{2}$ number density becomes higher if one increases ${\rm d}\varv/{\rm d}r$. Nevertheless, the derived H$_{2}$ number density only varies by a factor of two within a ${\rm d}\varv/{\rm d}r$ range of 1--10~\kms~pc$^{-1}$. 

\begin{figure*}[!htbp]
\centering
\includegraphics[width = 0.95 \textwidth]{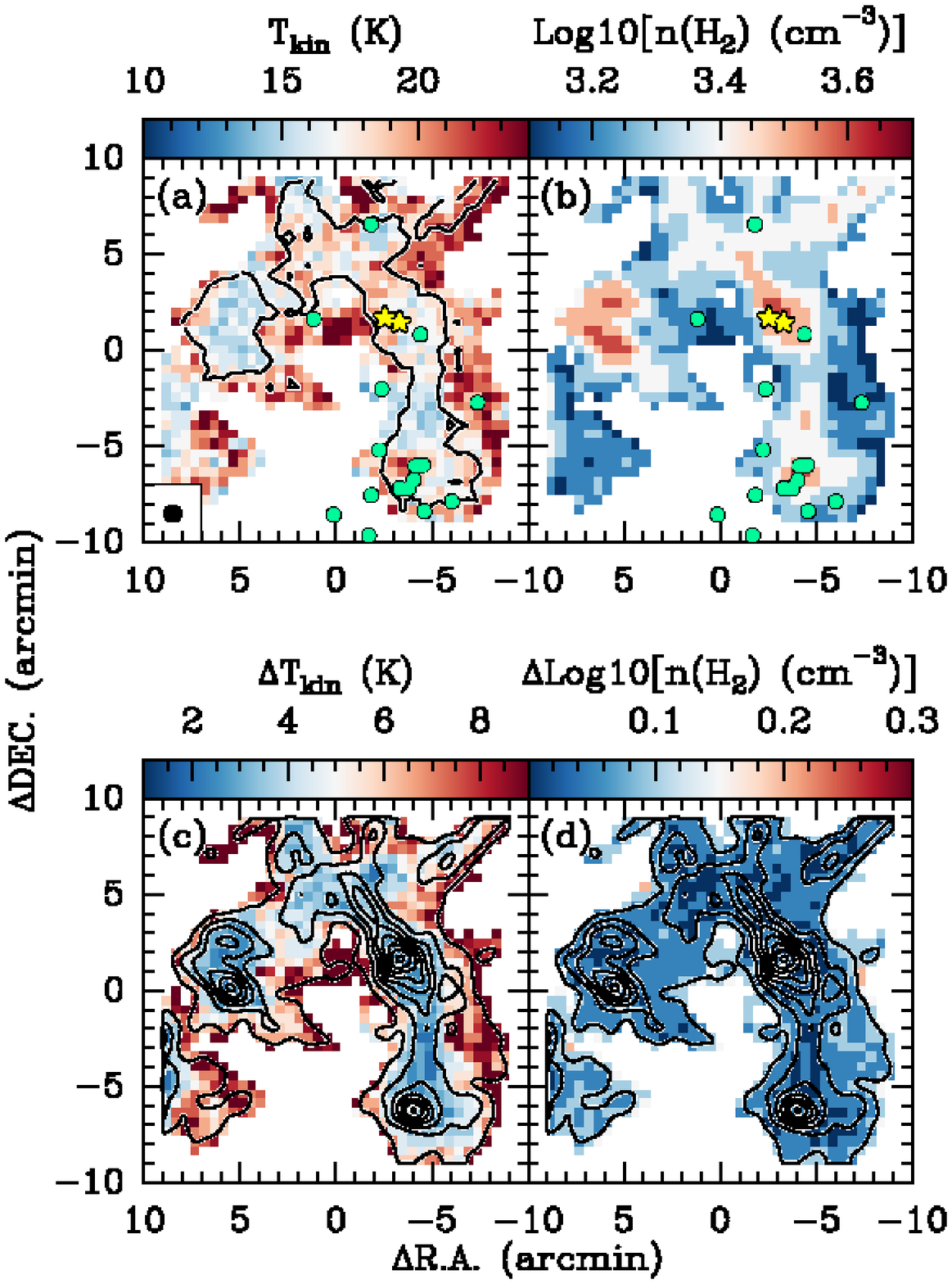}
\caption{{(a) Kinetic temperature map derived by the RADEX calculations. The black contour corresponds to a $T_{\rm kin}$/$\Delta T_{\rm kin}$ of 4, where $\Delta T_{\rm kin}$ is 1$\sigma$ uncertainty in $T_{\rm kin}$ (see Fig.~\ref{Fig:phy}c). (b) H$_{2}$ number density map derived by the RADEX calculations. In Figs.~\ref{Fig:phy}a--\ref{Fig:phy}b, the green filled circles represent the YSO candidates from \citet{2019MNRAS.484.1800S} and the two yellow pentagrams represent the two 22 GHz water maser positions. (c) Kinetic temperature uncertainty map overlaid with the $^{13}$CO (3--2) integrated intensity contours. (d) H$_{2}$ number density uncertainty map overlaid with the $^{13}$CO (3--2) integrated intensity contours. In Figs.~\ref{Fig:phy}c--\ref{Fig:phy}d, the contours start from 0.9 K~\kms\,(6$\sigma$) with a step of 0.9~K~\kms. In all panels, the (0, 0) offset corresponds to $\alpha_{\rm J2000}$=22$^{\rm h}$18$^{\rm m}$00$\rlap{.}^{\rm s}$44, $\delta_{\rm J2000}$=61\degree49\arcmin03$\rlap{.}^{\prime\prime}$7. }\label{Fig:phy}}
\end{figure*}

Following the method described above, we determined the kinetic temperature and H$_{2}$ number density of each pixel with the line ratios $R^{13}_{3-2/1-0}$ and $R^{13/12}_{3-2}$ at a linear resolution of 0.2~pc, that is, an angular resolution of 55\arcsec. Their distributions are shown in Fig.~\ref{Fig:phy}. Figure~\ref{Fig:phy}a shows that most of the derived kinetic temperatures lie in the range of 13--23 K with a median value of 18 K. These values are slightly higher than those ($\sim$10~K) obtained in other nearby dark clouds \citep[e.g.,][]{1989ApJS...71...89B,2017ApJ...843...63F,2017ApJ...850....3K}, but they are much lower than in OMC-1, parts of which are likely heated by radiation from the Trapezium Cluster \citep{2017ApJ...843...63F,2018A&A...609A..16T}. Although the pixels at the edges have higher kinetic temperature of $>$25 K, they have large uncertainties (see Fig.~\ref{Fig:phy}c) and are thus not included in our analysis. Higher kinetic temperatures are found within region C2 where star formation is more active than in the other regions. The high kinetic temperature could be due to feedback from YSOs. In contrast, H$_{2}$ number densities are better constrained to a range of 10$^{3}$--10$^{3.6}$ cm$^{-3}$. The derived H$_{2}$ number densities are lower than the critical density (2$\times 10^{4}$~cm$^{-3}$) of $^{13}$CO (3--2) \citep{2010ApJ...718.1062Y}, demonstrating that this transition is subthermal throughout the entire observed region. Regions C1, C2, and C3 have similar H$_{2}$ number densities of $\sim$10$^{3.6}$ cm$^{-3}$. Given that star formation has already begun in regions C1 and C2, we suggest that region C3 is the next potential star-forming region because of similar physical conditions in the three regions. We also note that self-absorption in $^{12}$CO (3--2) could affect the modeling results. In case of $^{12}$CO (3--2) self-absorption, the observed $R^{13/12}_{3-2}$ is supposed to be higher than the modeled values without considering $^{12}$CO (3--2) self-absorption. This will lead to an overestimate of the H$_{2}$ number density and an underestimate of kinetic temperature, indicating that region C3 tends to have H$_{2}$ number densities of $<$10$^{3.6}$ cm$^{-3}$ and kinetic temperatures of $\gtrsim$17 K.

%


\subsection{Revisiting the cloud-cloud collision scenario for L1188}\label{sec.cc}
In this section, we assess the link between the newly observed features and the cloud-cloud collision scenario initially proposed by \citet{2017ApJ...835L..14G}. In Sect.~\ref{res.mor}, we not only confirm the presence of the two orthogonal molecular clouds, but also discover complementary distributions between L1188a and L1188b, which represent emission in distinct velocity ranges (see Fig.~\ref{Fig:cpt}). Such complementary distributions are an expected outcome according to previous simulations of molecular cloud collisions \citep[e.g.,][]{1992PASJ...44..203H,2010MNRAS.405.1431A,2014ApJ...792...63T,2017ApJ...835..142T}. Furthermore, L1188 has no massive stars inside, excluding the possibility that such a complementary distribution is caused by the feedback of massive stars. Rather, it suggests a cloud-cloud collision in L1188 as its origin. Enhanced $^{13}$CO emission found in the contact regions of L1188a and L1188b (see the right panel of Fig.~\ref{Fig:cpt}) can result from a density enhancement caused by shocks expected during collisions. \citet{1976ApJ...209..466L} points out that colliding molecular clouds could provide heating in the interface, which can lead to self-absorption in optically thick lines due to temperature gradients. The observed widespread CO self-absorption (see Fig.~\ref{Fig:spec}) and high kinetic temperatures of $>$17~K (see discussions in Sect.~\ref{sec.phy}) further support our proposed scenario.

The two parallel filamentary structures F1 and F2 are nearly perpendicular to the elongated axis of L1188b, and each of them has at least two velocity components. We therefore propose that the formation of F1 and F2 can also be attributed to the collision during which molecular gas is stripped off or dragged out. In such a case, there should be shear between different velocity components, which would lead to the Kelvin-Helmholtz (KH) instability. The KH instability could result in the periodic variation of intensities found in F1, which is similar to the scenario proposed to explain the Taurus striations \citep{2016MNRAS.461.3918H}. For two incompressible fluids with densities $\rho_{1}$ and $\rho_{2}$, and velocities, $\upsilon_{1}$ and $\upsilon_{2}$, the maximum wavelength predicted by the KH instability \citep[e.g.,][]{1961hhs..book.....C} is
\begin{equation}\label{f.kh}
\lambda_{\rm KH, max} = \frac{2\pi}{g}\frac{\alpha_{1}\alpha_{2}}{\alpha_{1}-\alpha_{2}}(\upsilon_{1} -\upsilon_{2})^{2}\;,    
\end{equation}
where $\alpha_{1}$=$\rho_{1}/(\rho_{1}+\rho_{2})$, $\alpha_{2}$=$\rho_{2}/(\rho_{1}+\rho_{2})$, and $g$ is the acceleration. The velocity difference is taken to be 2.5~\kms\,according to Figs.~\ref{Fig:f1-pv}a--\ref{Fig:f1-pv}c. The density contrast ($\rho_{1}$/$\rho_{2}$) of two fluids in the interface still has to be discussed. The integrated intensity contrast is found about three between the two velocity components in Fig.~\ref{Fig:f1-pv}b. We assume the density contrast to be 3--10 at the initial contact stage. The acceleration is taken to be $g=\pi {\rm G}\mu m_{\rm H}N({\rm H_{2}})$, where G is the gravitational constant, $\mu$ is the mean molecular weight per hydrogen molecule which is assumed to be 2.8, $m_{\rm H}$ is the mass of the atomic hydrogen, and $N({\rm H_{2}})$ is the column density of the molecular hydrogen which is taken to be 2$\times 10^{21}$ cm$^{-2}$. This leads to a $\lambda_{\rm KH, max}$ range of  7--24 pc. These maximum wavelengths are much higher than the observed periodic scale of $~$0.8~pc in F1 (see Fig.~\ref{Fig:f1-pv}c). However, the observed regions differ from the ideal case \citep{1961hhs..book.....C}. \citet{1997ApJ...482..852H} incorporated thermodynamics into the KH instability by taking heating and CO cooling into account, and found that the perturbation wavelength is reduced to $\sim$0.9 pc for typical molecular regions (see their Sect.~3.4). The derived wavelength is similar to the observed periodic scale of $~$0.8~pc, indicating that such instability could occur in F1. 

The origin of the 1 pc-long shocked arc can also be attributed to the cloud-cloud collision. As shown in Sect.~\ref{sec.arc}, we identified two velocity components connecting each other with bridging features across the whole arc. Such a velocity structure has been predicted by previous simulations of cloud-cloud collisions \citep{2015MNRAS.454.1634H,2015MNRAS.450...10H}. Following the analyses for outflows \citep[e.g.,][]{2010ApJ...715.1170A}, we estimate the momentum and kinetic energy of the blueshifted high-velocity wing emission to be 1.4~M$_{\odot}$~\kms\,and 5.4$\times 10^{43}$~erg, respectively. According to previous theoretical predictions \citep[e.g.,][]{2018PASJ...70S..57W}, such momentum and kinetic energy can be easily created by cloud-cloud collisions. Alternatively, the derived momentum and kinetic energy also lie in the momentum and kinetic energy range of outflows from low-mass YSOs \citep{2010ApJ...715.1170A}. This indicates that even low-mass YSOs have the ability to inject the observed momentum and kinetic energy. However, the whole arc structure is nearly parallel to both the northeastern edge of L1188a and the centric concave shape of L1188b, and their linear scales are also similar. Such a morphology is more easily created by cloud-cloud collisions. Nevertheless, we cannot exclude the possibility that the 1-pc shocked arc is driven by the outflow from YSOs.

The star-forming activities in C1--C3 support the sequential star formation from west to east reported by \citet{2019MNRAS.484.1800S}. This can be alternatively explained by the cloud-cloud collision scenario, in which L1188b moved from west to east. This scenario would readily result in the observed filamentary YSO distribution which is nearly parallel to the long axis of L1188b (see Fig. 1). Based on the proposed scenario, star formation is expected in the northwest, especially in the compressed region C3 that has similar physical conditions to the known star-forming regions C1 and C2 (see Sect.~\ref{sec.phy}). \citet{2019MNRAS.484.1800S} suggest that star formation in L1188 started about 5 million years ago. Based on the assumption of the relative colliding speed of 5~\kms, \citet{2017ApJ...835L..14G} estimated that the collision occured about 1 Myr ago. If the periodic variation in F1 can be attributed to the KH instability, the characteristic growth time for KH instability \citep[$\tau_{\rm KH}=\frac{\lambda_{\rm KH, max}}{\upsilon_{1}-\upsilon_{2}}\frac{\rho_{1}+\rho_{2}}{(\rho_{1}\rho_{2})^{1/2}}$,][]{drazin1981hydrodynamic} could be an independent estimate for the cloud-cloud collision timescale. By using the initial H$_{2}$ number densities of two clouds before the collision in \citet{1997ApJ...482..852H}, we reach a lower limit of 0.7 Myr for $\tau_{\rm KH}$ because the currently observed periodic length of 0.8~pc is likely shorter than $\lambda_{\rm KH, max}$. Hydrodynamical simulations have shown that a timescale of 2--5 Myr is needed to generate an arc structure \citep{2014ApJ...792...63T}. The estimates point out that the collision might have happened about 0.7--5 Myr ago. This indicates that parts of YSOs in L1188 may have already formed before the collision, that is, parts of the star formation in L1188 are not caused by the collision. On the other hand, there are also stellar populations younger than 5 Myr. Therefore, the possibility of star formation triggered by the cloud-cloud collision should not be neglected.
  



\section{Summary}\label{Sec:sum}
In order to search for further observational evidence of cloud-cloud collisions in L1188, we performed multiple molecular line observations toward the 20\arcmin$\times$20\arcmin\, intersection region of the two nearly orthogonal filamentary molecular clouds. The main results are summarized as follows.
\begin{itemize}
\item[1.] Our $^{12}$CO (2--1) and $^{12}$CO/$^{13}$CO (3--2) observations provide a factor of approximately four better linear resolutions than previous CO observations. This leads to the discovery of a spatially complementary distribution between L1188a and L1188b. Enhanced $^{13}$CO emission and $^{12}$CO self-absorption is found toward their contacting regions. Furthermore, we found two filamentary structures F1 and F2, which are parallel to each other. Both of them have at least two velocity components that are connected with broad bridging features. At the most blueshifted side, we identified a 1 pc-long shocked arc ubiquitously showing $^{12}$CO line wings. 
  
\item[2.] In a 360\arcsec$\times$75\arcsec\,area of region C1 near the 1-pc shocked arc, we discover two positions showing 22 GHz water maser emission, which is the first maser detection in L1188. These masers are likely excited by the outflow from the Class I YSO WISEJ221729.87+615039.8. We also searched for thermal SiO emission and 95 GHz methanol maser emission, but none has been detected in the observed region. 

\item[3.] We performed a non-LTE analysis of line ratios toward the observed region. This suggests that L1188 is characterized by kinetic temperatures of 13--23~K and H$_{2}$ number densities of 10$^{3}$--10$^{3.6}$ cm$^{-3}$ at a linear resolution of 0.2 pc. High kinetic temperatures of $\gtrsim$20 K is found toward the most active star-forming region, which is likely due to feedback from YSOs. 
  
\item[4.] Compared with previous theoretical predictions and simulations, these new observational features can be readily explained by the scenario of cloud-cloud collisions. However, the feedback from low-mass young stellar objects may also make contributions to these observational features.
\end{itemize}

\section*{ACKNOWLEDGMENTS}\label{sec.ack}
We acknowledge the IRAM-30 m, JCMT, PMO-13.7 m, and Effelsberg-100 m staff for their assistance with our observations. This work was supported by the National Key R\&D Program of China under grant 2017YFA0402702, the National Natural Science Foundation of China (NSFC) under grant 11127903. X.D.T. acknowledges support by the NSFC under grant Nos. 11903070 and 11433008, and the Heaven Lake Hundred-Talent Program of Xinjiang Uygur Autonomous Region of China. C.H. acknowledges support by Chinese Academy of Sciences President's International Fellowship Initiative under grant No. 2019VMA0039. S.B.Z. was supported by the NSFC under grant 11803091. W.S.Z. was supported by the NSFC under grant 11673077. Y.W. acknowledge support from the European Research Council under the Horizon 2020 Framework Program via the ERC Consolidator Grant CSF-648505. The research leading to these results has received funding from the European Union’s Horizon 2020 research and innovation program under grant agreement No 730562 [RadioNet]. This work is based on observations with the 100-m telescope of the MPIfR (Max-Planck-Institut f{\"u}r Radioastronomie) at Effelsberg. The James Clerk Maxwell Telescope is operated by the East Asian Observatory on behalf of The National Astronomical Observatory of Japan; Academia Sinica Institute of Astronomy and Astrophysics; the Korea Astronomy and Space Science Institute; Center for Astronomical Mega-Science (as well as the National Key R\&D Program of China with No. 2017YFA0402700). Additional funding support is provided by the Science and Technology Facilities Council of the United Kingdom and participating universities in the United Kingdom and Canada. This publication makes use of data products from the Wide-field Infrared Survey Explorer, which is a joint project of the University of California, Los Angeles, and the Jet Propulsion Laboratory/California Institute of Technology, funded by the National Aeronautics and Space Administration. This research is based on observations with AKARI, a JAXA project with the participation of ESA. This research has made use of NASA's Astrophysics Data System. This work also made use of the AICer\footnote{https://github.com/shbzhang/aicer} plotting code, and Python libraries including Astropy\footnote{https://www.astropy.org/} \citep{2013A&A...558A..33A}, NumPy\footnote{https://www.numpy.org/} \citep{5725236}, SciPy\footnote{https://www.scipy.org/} \citep{jones2001scipy}, and Matplotlib\footnote{https://matplotlib.org/} \citep{Hunter:2007}. We thank the referee for helpful comments that improve the manuscript. Y.G. thanks C.-H. R. Chen and G.~X. Li for useful discussions.




\begin{appendix}
\section{The spectral energy distribution of WISEJ221729.87+615039.8}\label{sec.app}
\begin{figure}[!htbp]
\centering
\includegraphics[width = 0.45 \textwidth]{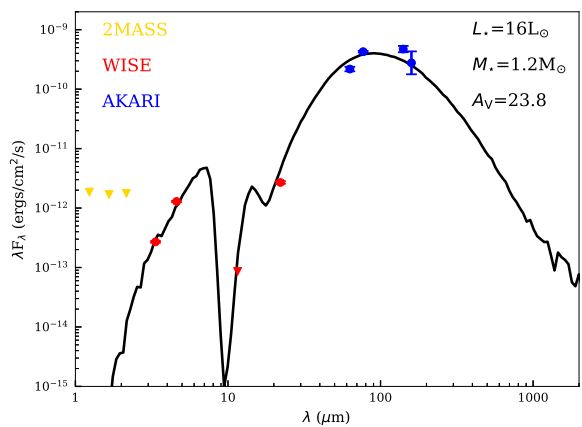}
\caption{{Spectral energy distribution of WISEJ221729.87+615039.8 from 1.2~$\mu$m to 160~$\mu$m. Dots represent measured data points, while triangles indicate upper limits. This SED was fit with YSO SED models \citep{2006ApJS..167..256R,2007ApJS..169..328R}. The solid black line indicates the best-fitting model.}\label{Fig:sed}}
\end{figure}

In order to study the nature of WISEJ221729.87+615039.8, we performed a spectral energy distribution (SED) fitting with the python SED fitter\footnote{https://sedfitter.readthedocs.io/en/stable/} \citep{2007ApJS..169..328R} that uses radiative transfer models of YSOs in \citet{2006ApJS..167..256R}. The modeled source distances are set to be within 700--1000 pc, while the modeled visual extinctions ($A_{\rm V}$) range from 0 to 40. Figure~\ref{Fig:sed} presents the SED fitting result, which employed infrared data from 2MASS \citep{2006AJ....131.1163S}, WISE \citep{2010AJ....140.1868W}, and AKARI \citep{2007PASJ...59S.389K}. This SED indicates that this YSO is a Class I object that is characterized by a 9.7~$\mu$m deep absorption of amorphous silicate in its circumstellar disk. From the best-fitting model, the distance is determined to be $\sim$788 pc and $A_{\rm V}$ is 23.8, confirming that this YSO is deeply embedded in L1188. The luminosity and stellar mass are estimated to be 16$L_{\odot}$ and 1.2$M_{\odot}$, suggesting a solar-type star. The inclination of the circumstellar disk is found to be about 57\degree, but this can be underestimated because we only have an upper limit at 12 $\mu$m. Because of the potential outflow origin of the H$_{2}$O masers (see Sect.~\ref{sec.maser}), WISEJ221729.87+615039.8 may have a disk-jet system.

\end{appendix}

\bibliographystyle{aa}
\bibliography{references}

\end{document}